\documentclass[11pt]{article}
\usepackage[preprint]{acl}
\usepackage{times}
\usepackage{latexsym}
\usepackage[T1]{fontenc}
\usepackage[utf8]{inputenc}
\usepackage{microtype}
\usepackage{inconsolata}
\usepackage{graphicx}
\usepackage{subcaption}
\usepackage{booktabs}
\usepackage{amsmath,amssymb,amsfonts}
\usepackage{multirow}
\usepackage{array}
\usepackage{tabularx}
\usepackage{enumitem}
\usepackage{url}
\usepackage{xcolor}
\usepackage{float}
\usepackage{placeins}
\usepackage{needspace}
\usepackage{hyperref}
\usepackage{cleveref}

\newcommand{\methodname}{\textsc{HyBIRD}}

\title{\methodname{}: Hyperbolic Bridge Retrieval and Diagnosis for Methodology Inspiration Retrieval}

\author{\normalfont
Yang Yang$^{1}$ \quad
Boyun Xu$^{1,3}$ \quad
Hao Fu$^{3}$ \quad
Jindong Li$^{1}$ \quad
Zining Zhong$^{1}$ \\
Bowen Tian$^{1}$ \quad
Jiemin Wu$^{1}$ \quad
Menglin Yang$^{1}$ \quad
Yutao Yue$^{1,2}$\thanks{Corresponding author: \texttt{yueyutao@hkust-gz.edu.cn}.} \\
$^{1}$The Hong Kong University of Science and Technology (Guangzhou), Guangzhou 511400, China \\
$^{2}$Institute of Deep Perception Technology, JITRI, Wuxi 214000, China \\
$^{3}$Shandong University, Weihai Campus, 180 Wenhua West Road, Weihai, 264209, Shandong, P.R. China
}

\begin{document}
\maketitle

\begin{abstract}
Methodology Inspiration Retrieval (MIR) asks a system to retrieve prior papers whose methods can inspire a new research proposal. Unlike general scientific retrieval, the central challenge is not topical similarity but whether a candidate paper provides concrete mechanisms that can instantiate an abstract methodological need. Existing MIR dense retrievers provide strong paper-level rankings, but the returned lists do not expose how proposal needs are bridged by retrieved methods, where evidence is weak, or which complementary snippets may help. We propose \methodname{}, a frozen-anchor framework that treats MIR as hyperbolic bridge retrieval and post-hoc method diagnosis. \methodname{} keeps a strong MIR dense retriever fixed, learns lightweight point, cone, and factorized hyperbolic bridge variants, and uses LLM-assisted method blocks for post-hoc explanation and evidence selection. On the MIR benchmark, the factorized bridge reaches 59.034 mAP while preserving the dense anchor's strong retrieval behavior. More importantly, \methodname{} converts ranked papers into inspectable query need profiles, factor coverage, maturity views, and complementary evidence bundles. The results suggest that hyperbolic geometry is most useful as calibrated structure over a dense anchor, rather than as a standalone replacement for dense retrieval.
\end{abstract}

\section{Introduction}

\begin{figure}[t]
\centering
\includegraphics[width=0.94\linewidth]{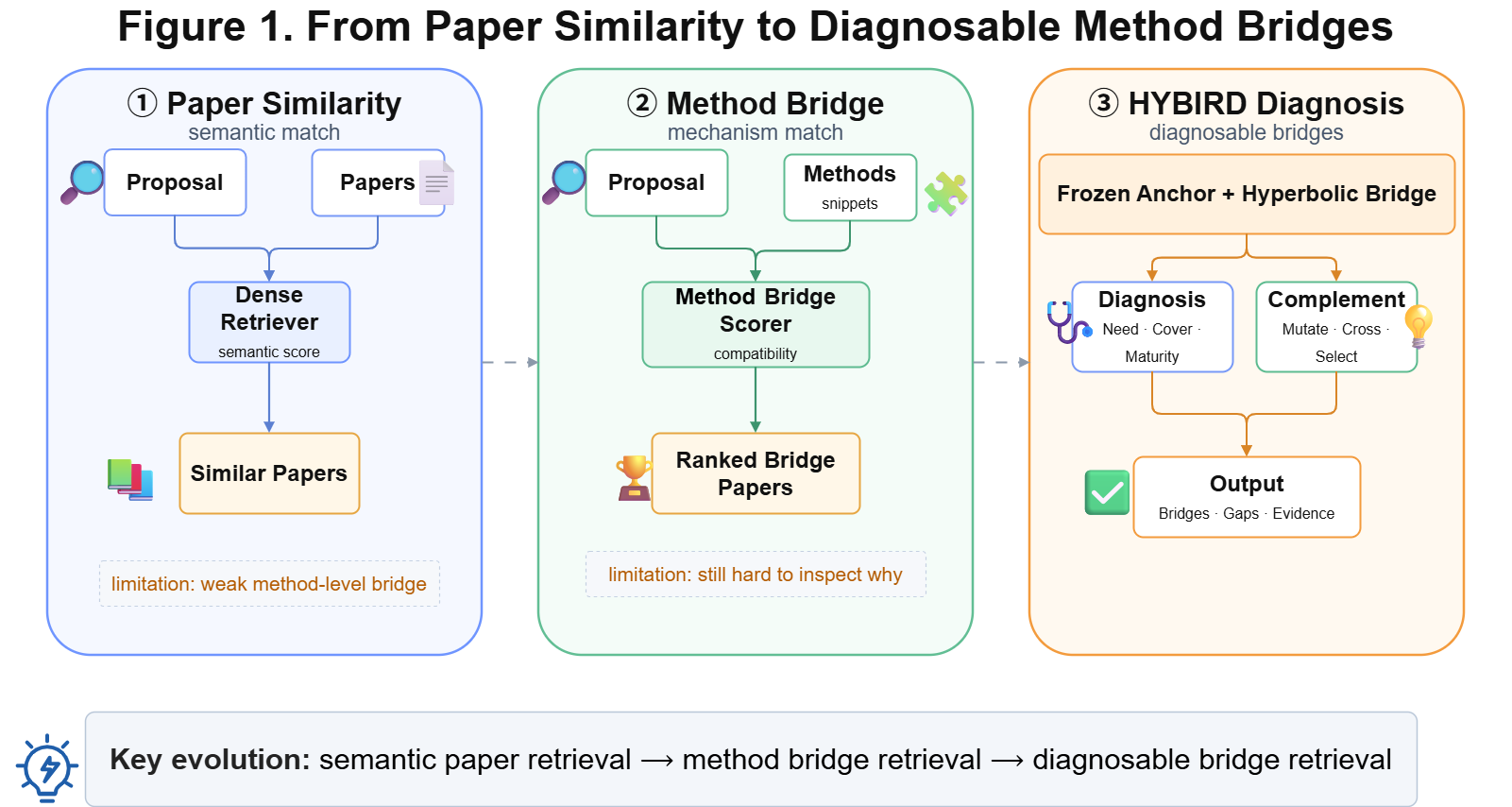}
\caption{Three retrieval views. Paper similarity retrieves topically similar documents, method bridge retrieval searches for mechanism compatibility, and \methodname{} adds post-hoc diagnosis over bridges, gaps, and complementary evidence.}
\label{fig:concept}
\end{figure}

Scientific inspiration often begins with retrieval: researchers must identify prior papers whose mechanisms can respond to a new proposal's needs. This step is increasingly important in literature-grounded ideation, research-agent, and AI-scientist-style workflows, where retrieved evidence grounds hypothesis generation, method planning, and experimental design \citep{swanson1986undiscovered,lewis2020rag,wang2024scimon,baek2025researchagent,asai2024openscholar,xu2026idea2story,lu2024ai,yamada2025aiscientistv2}. Human evaluations also show that LLM-generated ideas may appear novel while suffering from feasibility and diversity problems, making retrieval quality consequential \citep{si2024novelideas}.

However, topical relevance does not necessarily imply methodological applicability. A proposal about improving model compression may retrieve papers about sentence compression due to lexical overlap, while truly inspiring papers may instead involve teacher--student interaction, hidden-layer distillation, or representation transfer. Conversely, a paper from a different surface topic can still be methodologically useful if its mechanism can be adapted to the proposal's underspecified need. MIR is therefore closer to reasoning-intensive scientific search than to ordinary nearest-neighbor recommendation \citep{ajith2024litsearch,garikaparthi2025mir}.

\Cref{fig:concept} illustrates our target: MIR should return not only a ranked paper list, but an inspectable bridge from abstract methodological needs to concrete mechanism evidence, including covered factors, remaining gaps, and complementary snippets.

MIR established methodology inspiration retrieval as a distinct task \citep{garikaparthi2025mir}: given a proposal with problem and motivation but no method, retrieve prior papers whose methods can inspire a solution. MIR uses citation-intent supervision, including MultiCite-style contexts \citep{lauscher2022multicite}, to train methodology-aware dense retrievers. We keep the task, split, qrels, and evaluator unchanged, and ask how retrieved papers serve as methodological bridges.

We propose \methodname{}, a hyperbolic bridge framework for MIR and diagnosis. \methodname{} freezes a strong MIR dense retriever as an anchor, learns lightweight hyperbolic bridge scoring, and uses the same bridge view to identify factor-level needs, retrieved evidence, coverage gaps, maturity gaps, and complementary evidence bundles.

Our contributions are fourfold: we formulate MIR as methodological bridge retrieval; propose a frozen-anchor hyperbolic bridge scorer; analyze point, cone, and factorized variants; and convert retrieval traces into factor-level need, evidence, coverage-gap, maturity, and bundle-composition views.

\section{Related Work}
\subsection{Literature-grounded Scientific Discovery}
Literature-based discovery studies how disconnected findings suggest hypotheses \citep{swanson1986undiscovered}, while retrieval-augmented generation grounds language model outputs in evidence \citep{lewis2020rag}. Recent LLM systems use literature for ideation, synthesis, research-agent workflows, and AI-scientist-style automation \citep{wang2024scimon,baek2025researchagent,asai2024openscholar,xu2026idea2story,lu2024ai,yamada2025aiscientistv2}. These systems show the value of retrieved literature, but typically target broad relevance or complete idea generation. \methodname{} focuses on the narrower stage of ranking and diagnosing methodological inspirations.

\subsection{Scientific Retrieval with Citation and Method Signals}
Scientific retrieval often uses citation and scholarly text signals. SciBERT, Sentence-BERT, SPECTER, SciRepEval/SPECTER2, dense passage retrieval, LitSearch, S2ORC, and content-based citation recommendation provide strong foundations for scientific document representation, search, and recommendation \citep{beltagy2019scibert,reimers2019sbert,cohan2020specter,singh2023scirepeval,karpukhin2020dense,ajith2024litsearch,lo2020s2orc,bhagavatula2018content}. MIR moves this line from general paper similarity to methodology inspiration retrieval \citep{garikaparthi2025mir}, using citation-intent signals such as MultiCite \citep{lauscher2022multicite}. We keep the MIR protocol and study how a methodology-aware dense ranking can be geometrically calibrated and structurally interpreted.

\subsection{Scientific Evidence and Method Extraction}
Our method blocks and evidence bundles connect to scientific information extraction. SciERC and SciREX model scientific entities, relations, coreference, and document-level method/task/dataset/metric evidence \citep{luan2018scierc,jain2020scirex}. Citation-context resources such as MultiCite show that citation evidence is discourse-sensitive \citep{lauscher2022multicite}. We use method blocks post hoc to diagnose MIR outputs, not to build a general-purpose IE system or alter relevance labels.

\subsection{Geometric Inductive Biases for Hierarchy and Composition}
Hyperbolic geometry represents hierarchy with low distortion \citep{sarkar2011tree,nickel2017poincare,nickel2018lorentz,sala2018representation}. Order embeddings, hyperbolic cones, relation-specific cones, hyperbolic GCNs, and mixed-curvature product spaces further model asymmetric or heterogeneous structure \citep{vendrov2016order,ganea2018cones,bai2021cone,chami2019hyperbolic,gu2019learning}. Vision-language work such as MERU and PHyCLIP similarly uses hyperbolic or product geometry for hierarchy and composition \citep{radford2021clip,desai2023meru,yoshikawa2026phyclip}. \methodname{} adapts these biases to bridges from abstract proposal needs to concrete paper mechanisms.

\section{Method}
\subsection{Bridge Retrieval Setup}
Following MIR, each query proposal $q$ contains the research problem and motivation, but not the methodology. Given candidate papers $\mathcal{D}=\{d_1,\ldots,d_N\}$, MIR ranks papers by whether their methods can inspire the missing solution. We model each useful candidate as a bridge
\begin{equation}
  b(q,d,m): q \rightarrow d \rightarrow m,
\end{equation}
where $m$ is a method mechanism supplied by the paper. The unit of interpretation is therefore not only proposal--paper similarity, but proposal--paper--mechanism bridging.

\begin{figure*}[t]
\centering
\includegraphics[width=0.98\linewidth]{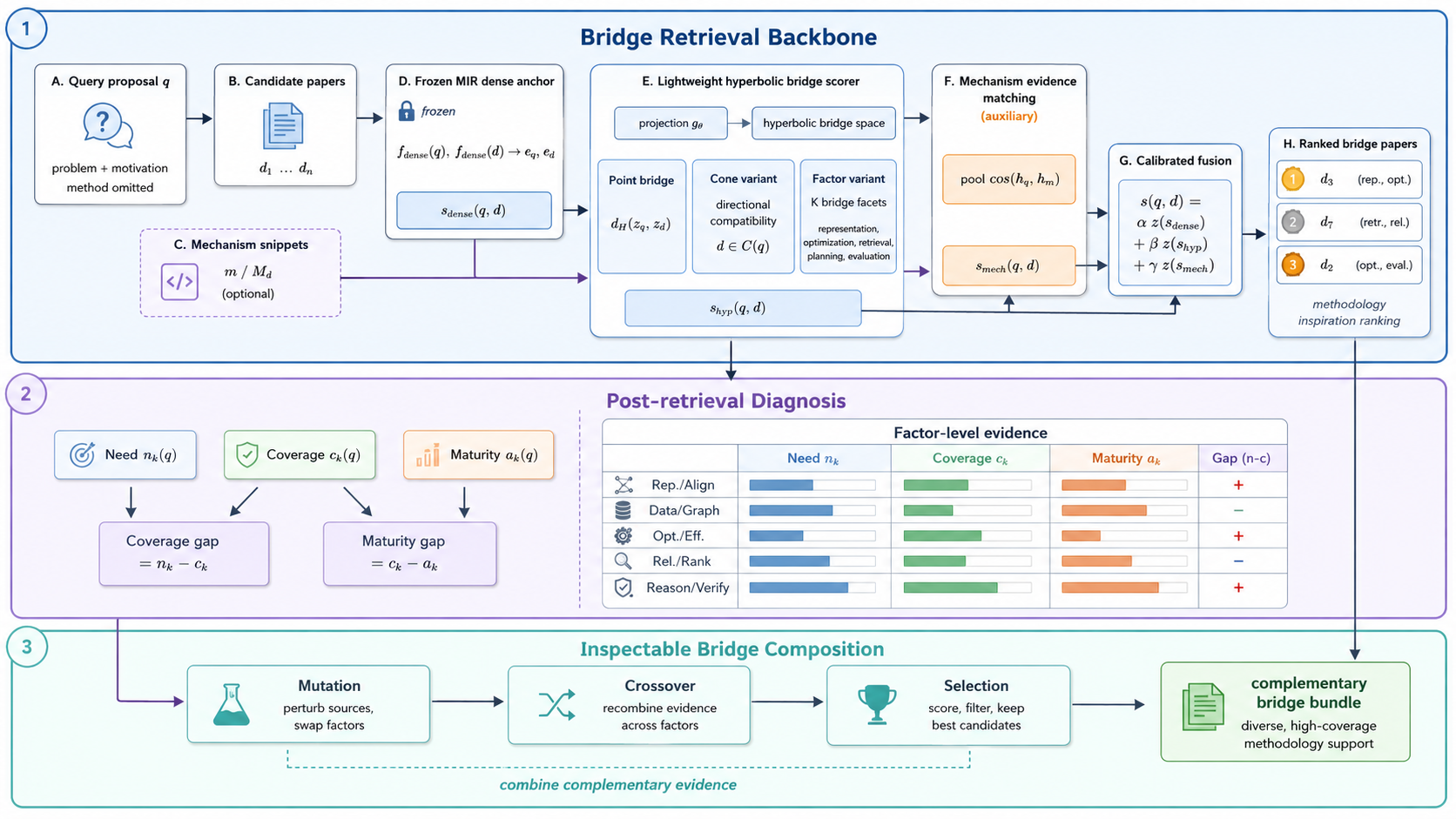}
\caption{Full \methodname{} workflow. The frozen MIR dense retriever supplies the anchor score. Hyperbolic point, cone, and factor variants add calibrated bridge structure. Method-block evidence is used for post-retrieval diagnosis and complementary bridge composition, not for changing qrels or the MIR evaluator.}
\label{fig:pipeline}
\end{figure*}

\subsection{Frozen Dense Anchor}
Let $f_{\mathrm{dense}}$ be the MIR fine-tuned dense retriever. We compute
\begin{equation}
  e_q=f_{\mathrm{dense}}(q),\quad e_d=f_{\mathrm{dense}}(d),
\end{equation}
and the dense anchor score $s_{\mathrm{dense}}(q,d)=\operatorname{sim}(e_q,e_d)$. We freeze this encoder in all main bridge variants. Freezing preserves the strong MIR representation and isolates the contribution of the bridge scorer. In \Cref{fig:pipeline}, this design is shown as the left-to-middle transition: dense paper similarity remains the retrieval backbone, while bridge geometry adds a second signal.

\subsection{Hyperbolic Point and Cone Bridges}
M1 maps dense representations into a hyperbolic bridge space:
\begin{equation}
  z_q=\operatorname{Exp}_0(g_{\theta}(e_q)),\quad z_d=\operatorname{Exp}_0(g_{\theta}(e_d)),
\end{equation}
and scores compatibility by negative hyperbolic distance:
\begin{equation}
  s_{\mathrm{hyp}}(q,d)=-d_{\mathbb{H}}(z_q,z_d).
\end{equation}
The fused retrieval score is calibrated with the dense anchor. M2 adds cone-style directionality: a paper is compatible when its representation lies in, or close to, the query-induced methodological cone $\mathcal{C}(q)$. This models MIR as an asymmetric relation from abstract need to concrete mechanism.

\subsection{Factorized Bridge and Method Blocks}
M3 factorizes the bridge into $K$ diagnostic facets. In our selected setting, $K=6$, corresponding to representation/alignment, classification/generation, data/graph/structure, optimization/efficiency, retrieval/ranking, and reasoning/verification/repair.

\paragraph{Factor ontology and weak supervision.}
The six factor labels are weak diagnostic roles rather than manually exhaustive annotations. We use a compact MIR-oriented ontology with six roles: representation/alignment, classification/generation, data/graph/structure, optimization/efficiency, retrieval/ranking, and reasoning/verification/repair. These roles are bootstrapped from query and method semantics, citation-intent supervision, and lightweight role-matching heuristics over proposal text and method snippets. The resulting factor assignments are soft and partially supervised: they serve as diagnostic structure for bridge analysis, not as gold semantic classes. We select $K=6$ because it yields the best factor quality and interpretability on the validation sweep.

M3 is trained with retrieval, role alignment, sparse activation, factor diversity, and cone losses:
\begin{equation}
\begin{aligned}
L ={}& L_{\mathrm{ret}}
 + \lambda_{\mathrm{role}} L_{\mathrm{role}}
 + \lambda_{\mathrm{sparse}} L_{\mathrm{sparse}} \\
&+ \lambda_{\mathrm{div}} L_{\mathrm{div}}
 + \lambda_{\mathrm{cone}} L_{\mathrm{cone}} .
\end{aligned}
\end{equation}
We set $\lambda_{\mathrm{role}}=0.10$, $\lambda_{\mathrm{sparse}}=0.10$, $\lambda_{\mathrm{div}}=0.05$, and $\lambda_{\mathrm{cone}}=0.01$.

For diagnosis, we use LLM-assisted method blocks extracted from abstracts. In our implementation, the extraction backend is DeepSeek-v4pro. Each block contains a method name, mechanism, target problem, short label, evidence sentence, and confidence. The blocks are used for M4/M5 evidence and visualization only; they are not new relevance labels and do not alter MIR qrels.

\subsection{M4: Query-conditioned Gap Diagnosis}
\label{sec:gap_m5}
For each query $q$ and factor $k$, M4 computes three diagnostic quantities. The query need score $n_k(q)$ is the normalized factor-keyword activation over the proposal text. For a retrieved method block $u$, we compute its factor score $s_k(u)$ using the same factor lexicon and compute lexical overlap $\operatorname{sim}(q,u)$ between the query and the method block. Retrieved coverage is the best supported evidence among the top-$K_r$ retrieved method blocks:
\begin{equation}
c_k(q)=\max_{u\in R(q)}
\left(0.65\,s_k(u)+0.35\,\operatorname{sim}(q,u)\right),
\end{equation}
where $R(q)$ denotes the retrieved method blocks used for diagnosis. We define maturity $a_k(q)$ as the global percentile of the selected best block's factor score among all DeepSeek-v4pro method blocks for factor $k$:
\begin{equation}
\begin{aligned}
a_k(q)&=
\operatorname{Percentile}_{v\in\mathcal{U}}
\left(s_k(u^\star); s_k(v)\right),\\
u^\star&=\arg\max_{u\in R(q)}
\left(0.65\,s_k(u)+0.35\,\operatorname{sim}(q,u)\right).
\end{aligned}
\end{equation}

These quantities separate three diagnostic views. The need score $n_k(q)$ indicates which method factor the proposal asks for; coverage $c_k(q)$ checks whether the retrieved method blocks contain evidence for that factor; and maturity $a_k(q)$ asks whether the best retrieved evidence is specific relative to the global method-block distribution. Thus, coverage measures whether evidence exists for a needed factor, while maturity measures how strong or mature that evidence is among all extracted method blocks.

We assign a discrete diagnosis label for each factor using the following rule:
\begingroup
\small
\begin{equation}
\operatorname{Diag}_k(q)=
\begin{cases}
\text{non-core}, & n_k(q)<0.05,\\
\text{strong}, & c_k(q)\ge0.8\,n_k(q),\\
\text{partial}, & 0.5\,n_k(q)\le c_k(q)<0.8\,n_k(q),\\
\text{gap}, & c_k(q)<0.5\,n_k(q).
\end{cases}
\end{equation}
\endgroup
Intuitively, a non-core factor is not required by the query; a strong factor is needed and covered by retrieved method evidence; a partial factor has some support; and a gap factor is needed but weakly covered. Coverage gaps and maturity gaps are computed as
\begin{align}
\mathrm{CoverageGap}_k(q)&=n_k(q)-c_k(q),\\
\mathrm{MaturityGap}_k(q)&=c_k(q)-a_k(q).
\end{align}

\subsection{M5: Complementary Evidence Bundles}
M5 then selects complementary evidence bundles with an internal fitness function combining relevance, evidence strength, coverage, coherence, and redundancy penalty. For a candidate bundle $B=\{b_1,\ldots,b_m\}$ and query $q$, we use
\begin{equation}
\begin{aligned}
\mathrm{Fit}(B;q)={}&0.30\,\mathrm{Rel}(B,q)
 +0.25\,\mathrm{Evid}(B)\\
&+0.25\,\mathrm{Cov}(B,q)
 +0.15\,\mathrm{Coh}(B)\\
&-0.05\,\mathrm{Red}(B).
\end{aligned}
\end{equation}
Here $\mathrm{Rel}$ is query--method relevance, $\mathrm{Evid}$ is method-block evidence strength, $\mathrm{Cov}$ measures coverage of active M4 factor gaps, $\mathrm{Coh}$ measures within-bundle coherence, and $\mathrm{Red}$ penalizes redundant blocks. All terms are normalized before aggregation. This score is used only to choose inspectable bridge bundles; it is not mAP and not a retrieval metric.

\section{Experiments}
\subsection{Setup}
We evaluate on the MIR benchmark with the original data, chronological split, qrels, and evaluator preserved. Retrieval metrics are Recall@3, Recall@5, and mAP. Factor diagnostics use factor quality, normalized entropy, max factor share, role accuracy, and snippet coherence. M4/M5 do not train a new retriever, rerank benchmark candidates, alter qrels, or change metric computation.
We apply the M4 diagnosis rule from \Cref{sec:gap_m5} to all released test query IDs and select representative cases for visualization.

We follow the original MIR chronological splits with 704/86/139 original proposals for train/dev/test. In the released CSV files, these correspond to 800/102/157 generated query IDs because multiple generated query instances may map to the same original proposal. All benchmark retrieval results follow the original MIR split and qrels; the generated query ID count is reported only to clarify the released-file statistics.

\begin{table*}[t]
\centering
\small
\setlength{\tabcolsep}{4pt}
\begin{tabular}{lrrrrrr}
\toprule
Split & Original Proposals & Generated Query IDs & Pairs & Candidates & Positives & Pos./Generated Query \\
\midrule
Train & 704 & 800 & 7551 & 1232 & 2543 & 3.18 \\
Dev & 86 & 102 & 1004 & 179 & 328 & 3.22 \\
Test & 139 & 157 & 1698 & 284 & 660 & 4.20 \\
\bottomrule
\end{tabular}
\caption{MIR dataset statistics under the original chronological splits. Original proposals follow the MIR paper statistics, while generated query IDs are counted from the released CSV files; multiple generated query IDs may map to the same original proposal. Candidate counts are unique cited-paper identifiers within each split. The DeepSeek-v4pro method-block cache covers 4,799 documents and contains 7,717 method blocks.}
\label{tab:dataset_stats}
\end{table*}

\Cref{tab:dataset_stats} reports the dataset statistics with both original-proposal counts and generated query IDs. Each generated query ID is associated with about 9--11 candidate rows in the released chronological files, and positives are defined by the released inspirational labels. We additionally use a method-block cache covering 4,799 documents, which is larger than the direct train/dev/test candidate union because it supports the extended M4/M5 evidence scan.

\paragraph{Artifacts.}
The artifacts used in this work are the MIR benchmark \citep{garikaparthi2025mir}, MultiCite-derived citation-intent supervision \citep{lauscher2022multicite}, the MIR dense retriever initialized from Stella 1.5B FT, and LLM-assisted method blocks extracted with DeepSeek-v4pro. The released MIR README describes the benchmark as an extension of MultiCite and provides the chronological splits used in our experiments. We use these publicly released research artifacts for research evaluation, follow their released terms and usage conditions, do not redistribute the original raw data, and keep the original MIR split, qrels, and evaluator unchanged.

\subsection{Main Retrieval Results}
\Cref{tab:main} reports main retrieval results. We include the MIR paper's reported Table 2 Stella\_1.5B FT extended-corpus result as a reference point, followed by our local dense-anchor recomputation and the three \methodname{} bridge variants. The dense anchor remains strong. HyBIRD variants improve Recall@5 and mAP relative to the local anchor, but do not uniformly improve Recall@3. This suggests that bridge calibration mainly improves ranking depth rather than simply pushing every relevant paper into the first positions.

\begin{table}[t]
\centering
\small
\setlength{\tabcolsep}{3pt}
\begin{tabular}{llccc}
\toprule
Model & Description & R@3 & R@5 & mAP \\
\midrule
MIR Table 2 & Stella\_1.5B FT extended & 60.110 & 65.320 & 56.890 \\
M0 & Local MIR dense anchor & 61.032 & 64.892 & 58.058 \\
M1 & HyBIRD-Point & 59.849 & 67.796 & 58.848 \\
M2 & HyBIRD-Cone & 59.688 & 67.581 & 58.846 \\
M3 & HyBIRD-Factor & 59.688 & 67.581 & 59.034 \\
\bottomrule
\end{tabular}
\caption{Main retrieval results on MIR. The first row gives the MIR paper's reported Stella\_1.5B FT extended-corpus result; all \methodname{} variants are built on the local frozen dense anchor.}
\label{tab:main}
\end{table}

The result pattern is intentionally not framed as ``hyperbolic geometry beats dense retrieval.'' Instead, \methodname{} is a calibrated bridge over a strong dense ordering. The local dense-anchor recomputation is close to the MIR Table 2 reference, supporting that the retained split and evaluator produce comparable behavior. The R@5 gain indicates that the bridge can recover additional methodologically relevant papers slightly deeper in the ranking, while the R@3 tradeoff shows that the dense anchor already captures many obvious top-ranked positives. In a literature-exploration setting, this is still useful because researchers often inspect more than the first three candidates.

\subsection{Ablation Study}
\Cref{tab:ablation} shows that hyperbolic-only scoring is much weaker than the dense anchor. This supports our central interpretation: hyperbolic structure is useful as calibrated bridge structure over a strong dense retriever, not as a standalone replacement.

\begin{table}[t]
\centering
\small
\begin{tabular}{lccc}
\toprule
Variant & R@3 & R@5 & mAP \\
\midrule
MIR baseline & 61.032 & 64.892 & 58.058 \\
MIR-anchor control & 60.387 & 64.570 & 57.930 \\
Hyperbolic-only & 55.065 & 61.989 & 53.402 \\
HyBIRD-Point / M1 & 59.849 & 67.796 & 58.848 \\
HyBIRD-Cone / M2 & 59.688 & 67.581 & 58.846 \\
HyBIRD-Factor / M3 & 59.688 & 67.581 & 59.034 \\
\bottomrule
\end{tabular}
\caption{Ablation study. The dense anchor is essential; factorized hyperbolic structure is most useful as calibrated correction and diagnosis.}
\label{tab:ablation}
\end{table}

The ablation separates three effects. The MIR-anchor control measures the score behavior without hyperbolic geometry. The hyperbolic-only row removes the dense anchor and drops sharply, confirming that geometry alone does not provide enough corpus-level grounding. The point, cone, and factor rows show that bridge structure is best treated as a lightweight correction and analysis layer.

\subsection{Factor Selection and Semantics}
\begin{figure}[!htbp]
\centering
\includegraphics[width=0.98\linewidth]{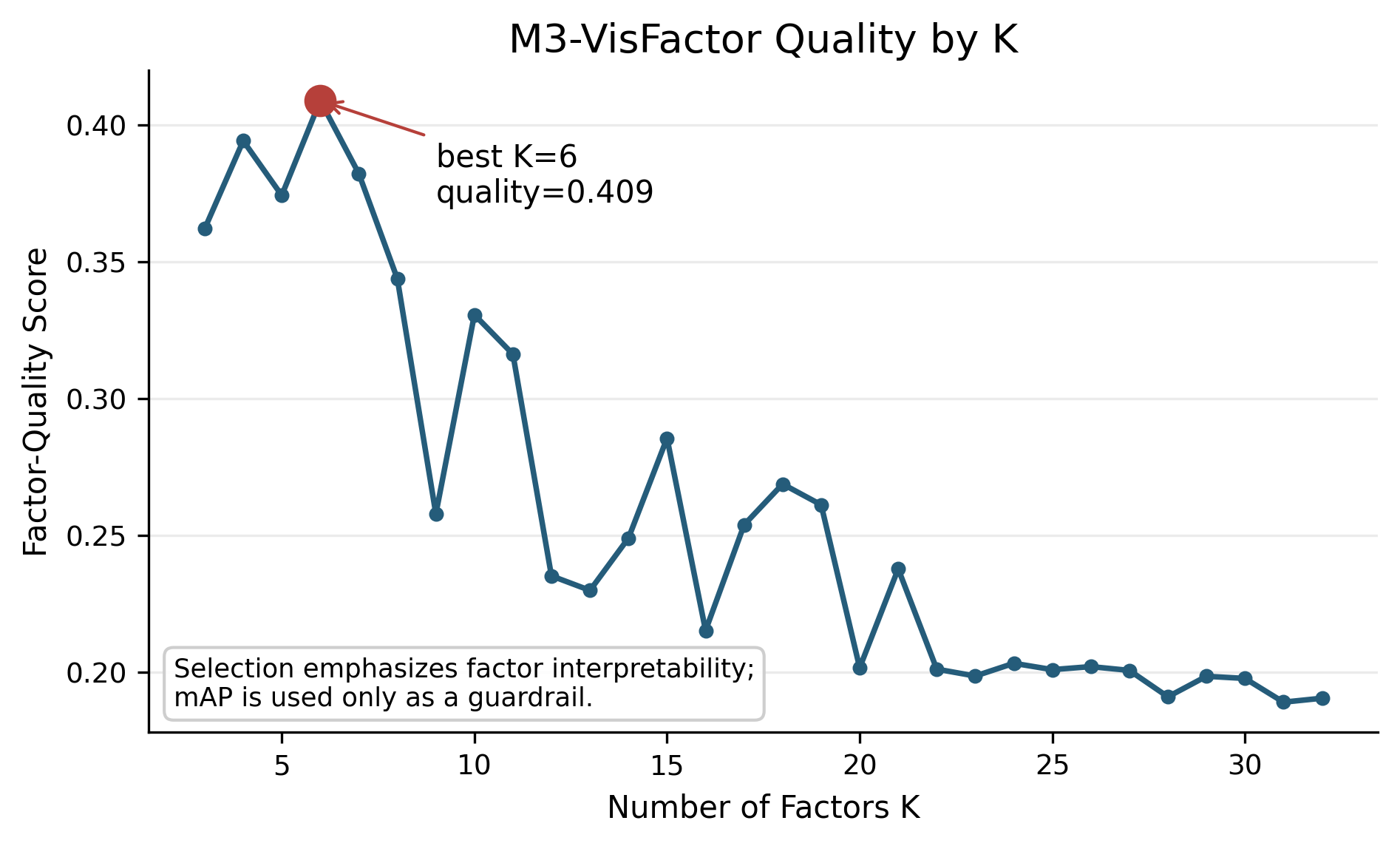}
\caption{Factor-quality score across $K$ values. The selected $K=6$ setting gives the best diagnostic quality among the sweep.}
\label{fig:k_quality}
\end{figure}

\begin{figure}[!htbp]
\centering
\includegraphics[width=0.98\linewidth]{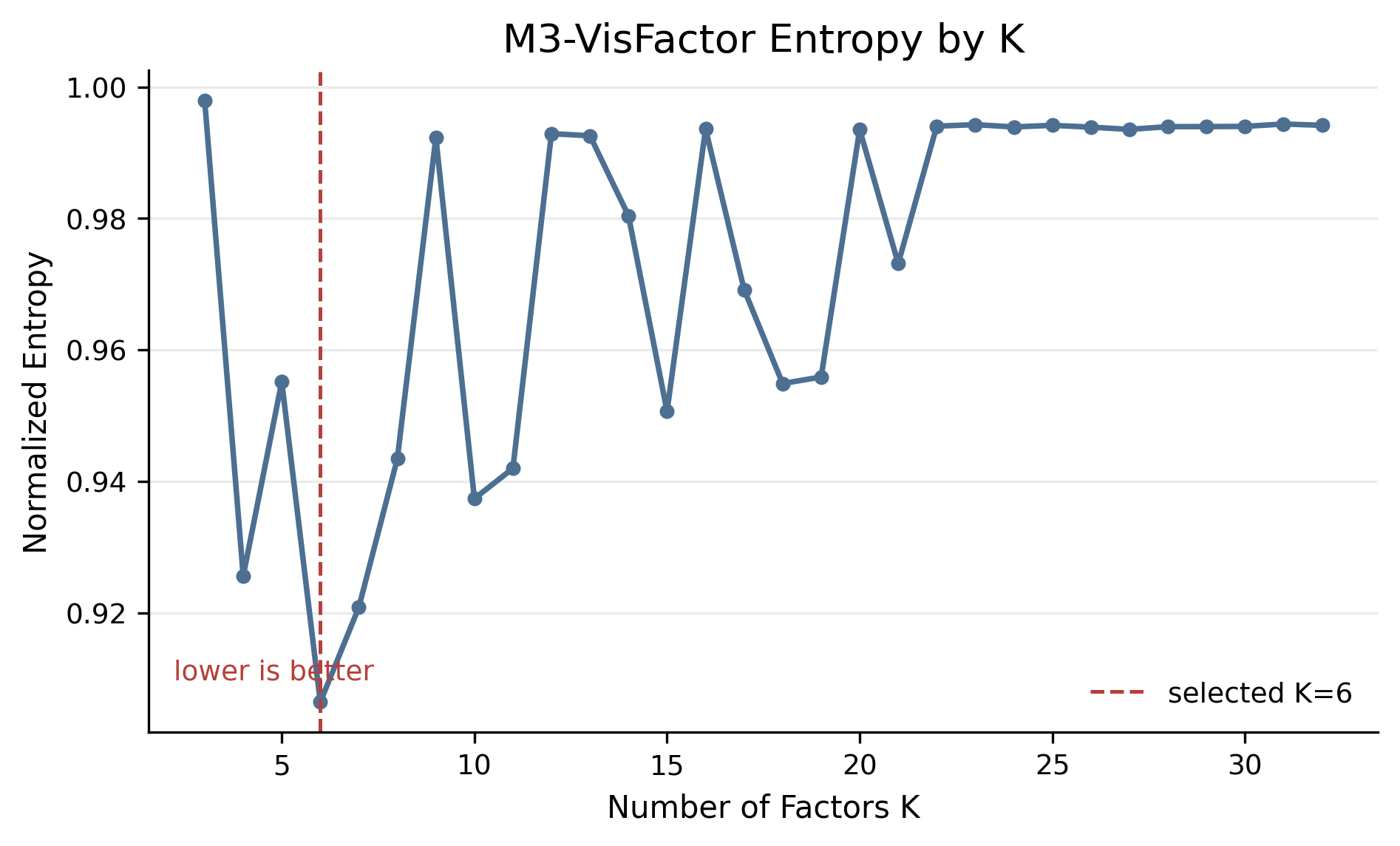}
\caption{Normalized factor entropy across $K$ values. Larger factor counts become more uniformly activated and less useful as compact diagnostic roles.}
\label{fig:k_entropy}
\end{figure}

Factorization is useful only if the factors are inspectable. We therefore select $K$ using factor quality and diagnostic behavior rather than mAP alone. \Cref{fig:k_quality} and \Cref{fig:k_entropy} should be read together: the selected $K=6$ setting has the strongest factor-quality score, while larger $K$ values become increasingly diffuse.

\begin{table*}[t]
\centering
\small
\begin{tabular}{rrrrrrrl}
\toprule
$K$ & Quality & Dev mAP & Entropy & Max share & Role acc. & Coherence & Selected \\
\midrule
4 & 0.394 & 59.715 & 0.926 & 0.378 & 0.167 & 0.471 & \\
5 & 0.374 & 59.690 & 0.955 & 0.296 & 0.167 & 0.414 & \\
6 & 0.409 & 59.690 & 0.906 & 0.309 & 0.202 & 0.472 & yes \\
8 & 0.344 & 59.691 & 0.943 & 0.234 & 0.197 & 0.410 & \\
10 & 0.331 & 59.690 & 0.937 & 0.221 & 0.169 & 0.418 & \\
12 & 0.235 & 58.901 & 0.993 & 0.112 & 0.167 & 0.382 & \\
16 & 0.215 & 58.896 & 0.994 & 0.089 & 0.153 & 0.379 & \\
20 & 0.201 & 58.897 & 0.994 & 0.072 & 0.140 & 0.383 & \\
24 & 0.203 & 58.896 & 0.994 & 0.062 & 0.178 & 0.378 & \\
32 & 0.190 & 58.895 & 0.994 & 0.047 & 0.170 & 0.371 & \\
\bottomrule
\end{tabular}
\caption{M3 factor-quality sweep. $K=6$ is selected for diagnostic quality and interpretability rather than mAP alone.}
\label{tab:k}
\end{table*}

\begin{figure}[H]
\centering
\includegraphics[width=0.98\linewidth]{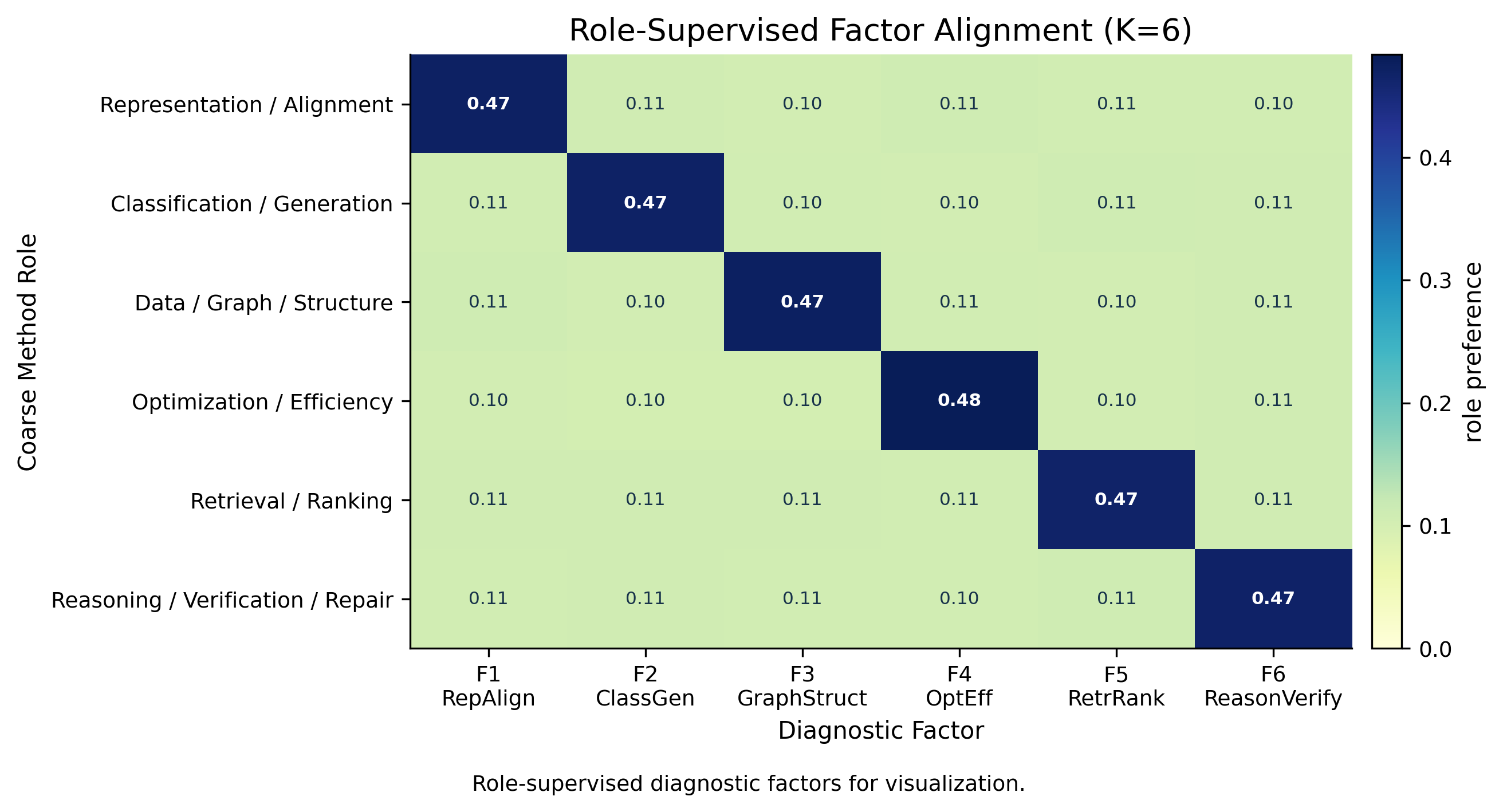}
\caption{Role-factor alignment heatmap for the selected $K=6$ M3-Factor backbone. Each factor is interpreted post hoc using role supervision and top method snippets.}
\label{fig:role_alignment}
\end{figure}

\Cref{fig:role_alignment} provides the qualitative counterpart to \Cref{tab:k}. The heatmap does not prove that factors are fully unsupervised method categories. It shows that the selected role-supervised backbone can organize evidence into recurring diagnostic roles such as representation/alignment, classification/generation, and retrieval/ranking.

\section{Analysis and Case Study}
\begin{figure*}[!t]
\centering
\includegraphics[width=0.88\textwidth]{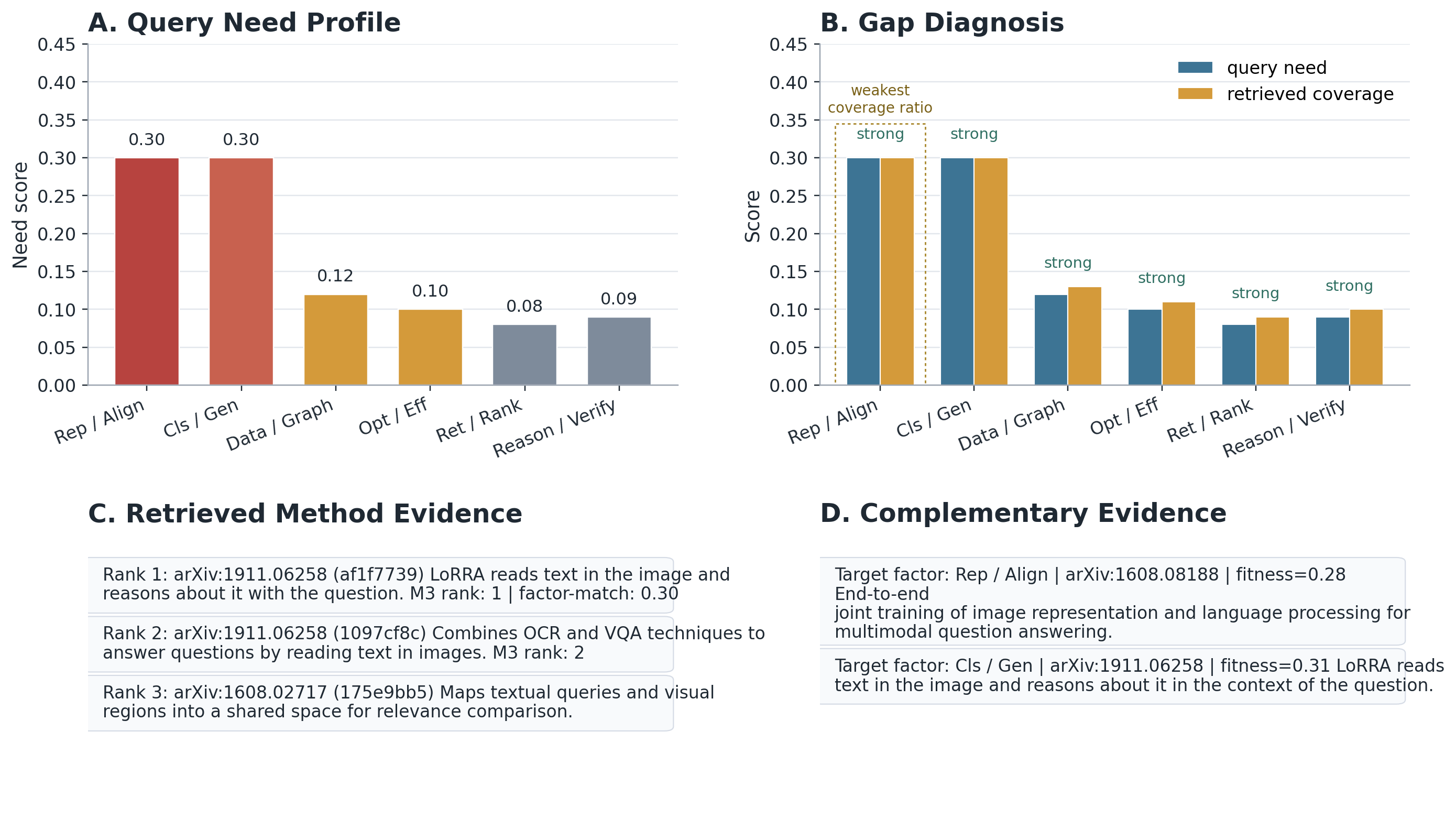}
\caption{M4/M5 case study. The four panels show query need, gap diagnosis, retrieved method evidence, and complementary evidence. The figure is intended to be read as an explanation path rather than as an additional ranking metric.}
\label{fig:case}
\end{figure*}

The M4/M5 layer is designed to answer questions that a ranked paper list cannot answer: which method roles does the query need, which roles are covered by retrieved papers, and which snippets can complement weak factors? \Cref{fig:case} shows a representative multimodal question answering case. The panels separate query need, gap diagnosis, retrieved method evidence, and complementary evidence, so the reader can trace the explanation from the query profile to the recommended bridge bundle.

In the shown case, the query strongly activates representation/alignment and classification/generation factors. The retrieved evidence covers part of this need, while complementary evidence proposes snippets that strengthen the missing or weak factors. This is the intended user-facing behavior of \methodname{}: after a dense retriever returns candidate papers, the diagnostic layer explains which methodological bridge each paper contributes and which evidence would make the bridge more complete.

\begin{center}
\centering
\includegraphics[width=0.72\linewidth]{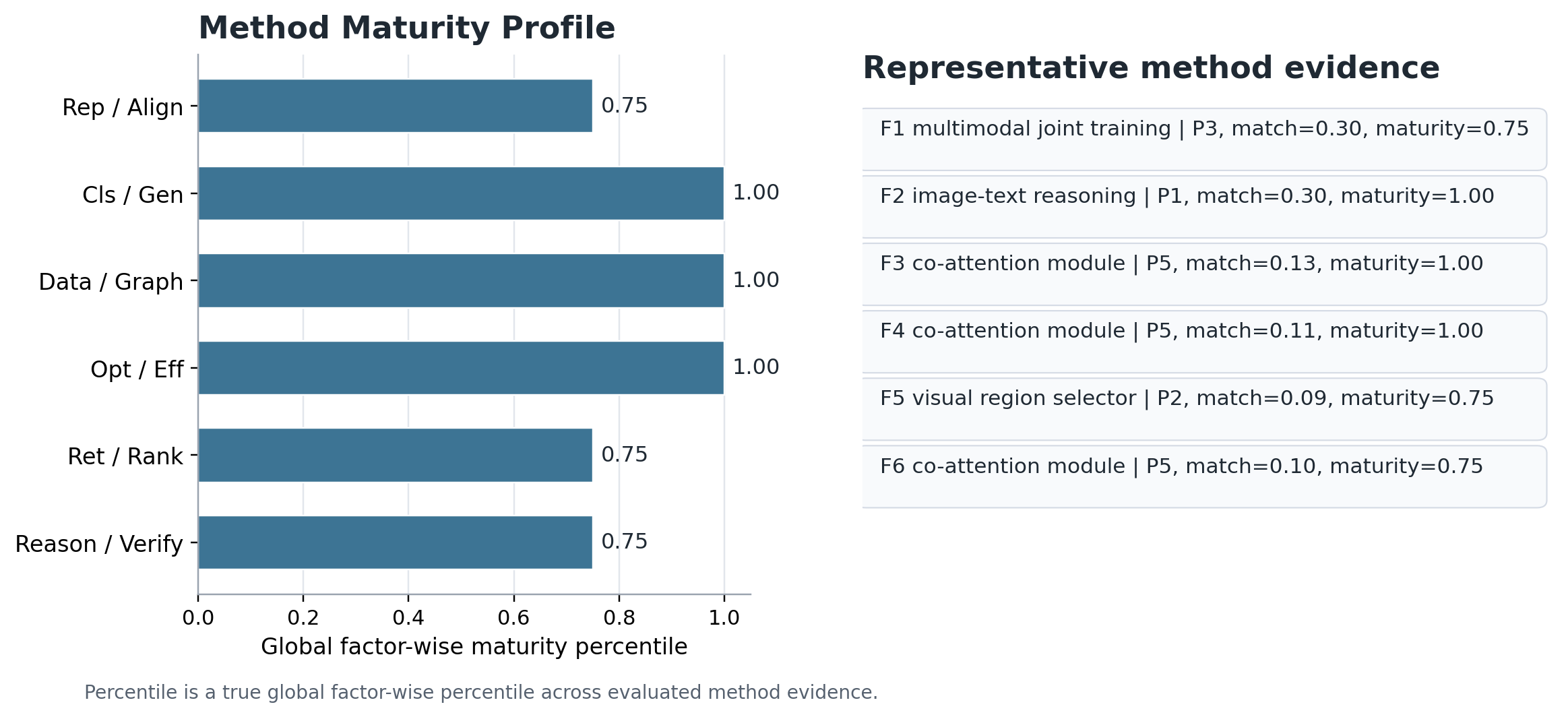}
\captionof{figure}{Method maturity profile estimated from retrieved method evidence. Maturity is separated from coverage so that generic evidence is not treated as fully adequate support.}
\label{fig:maturity}
\end{center}

\Cref{fig:maturity} gives a more compact maturity view for the same diagnosis family. Coverage indicates whether some evidence exists for a factor; maturity asks whether that evidence is specific and usable. This distinction is important for MIR because a paper can mention a relevant method family without providing the concrete mechanism needed for proposal development.

\subsection{M5 Evidence-Bundle Comparison}
After the case-level explanation, we aggregate the M5 evidence-bundle fitness scores in \Cref{tab:m5}. The proposed factor-gap method outperforms no-hyperbolic and random-block alternatives under the internal fitness function defined in \Cref{sec:gap_m5}. This result supports factor-guided evidence selection, but it should not be interpreted as retrieval accuracy or scientific correctness.

\begin{table}[t]
\centering
\small
\begin{tabular}{lrrr}
\toprule
Method & Mean fitness & Std & Median \\
\midrule
Ours & 0.590 & 0.045 & 0.607 \\
No Hyperbolic & 0.494 & 0.076 & 0.526 \\
Random Method Blocks & 0.475 & 0.083 & 0.501 \\
\bottomrule
\end{tabular}
\caption{M5 evidence-bundle fitness comparison. Fitness is an internal evidence score, not mAP.}
\label{tab:m5}
\end{table}

\subsection{Bundle Construction}
For each query, we keep the retrieved top-3 method blocks as the origin bundle. Mutation augments this origin bundle with factor-conditioned complementary blocks selected from LLM-extracted method evidence; in our current implementation, mutation appends new blocks and does not delete origin blocks. Crossover constructs a retrieved-evidence bundle by selecting one best retrieved block for each active factor, without reusing the same document--block pair. Mutation+crossover then combines the crossover bundle with the same complementary blocks used for mutation. These bundle paths are used only for post-hoc evidence selection and diagnosis, not for MIR ranking or qrel modification.

\FloatBarrier
\section{Conclusion}
We introduced \methodname{}, a frozen-anchor framework for methodology inspiration retrieval with hyperbolic bridge structure and post-hoc method diagnosis. The results show that hyperbolic-only retrieval is insufficient, but dense-plus-hyperbolic bridge variants remain competitive and provide useful interpretability. The selected M3-Factor backbone enables query need profiles, gap maps, and complementary evidence bundles through M4/M5 diagnosis. \methodname{} therefore reframes MIR from a paper-listing problem into a diagnosable method bridge problem while preserving a conservative distinction between benchmark retrieval and explanatory evidence.

\section*{Limitations}
\methodname{} does not claim that M4/M5 improve MIR benchmark ranking. These modules are post-hoc diagnosis and complementary evidence tools. The factor labels are role-supervised and post-hoc interpreted, so they should be treated as diagnostic categories rather than fully unsupervised discovered method taxonomies. Method blocks are extracted from abstracts rather than full papers, which can miss implementation details. DeepSeek-v4pro extraction may introduce errors and should be checked by experts before any suggested method is reused. Finally, M5 fitness is an internal score for evidence-bundle selection and does not prove downstream scientific usefulness without a user study.

\section*{Ethics Statement}
\methodname{} is intended to support literature exploration and proposal development. It should not replace expert reading of the original papers. LLM-extracted method blocks may be incomplete or incorrect, so users should verify all evidence against source papers. The system may also inherit publication, citation, and dataset biases from the MIR corpus. Retrieved and recommended methods should be attributed correctly and should not be treated as generated original contributions.

\bibliography{references}

\appendix
\onecolumn
\raggedbottom
\setlength{\textfloatsep}{8pt plus 2pt minus 2pt}
\setlength{\floatsep}{8pt plus 2pt minus 2pt}
\setlength{\intextsep}{8pt plus 2pt minus 2pt}
\setlength{\abovecaptionskip}{3pt}
\setlength{\belowcaptionskip}{2pt}
\setlength{\tabcolsep}{5pt}
\renewcommand{\arraystretch}{1.12}

\section{Artifact and Reproducibility Details}
\begin{center}
\centering
\small
\begin{tabularx}{0.96\textwidth}{p{0.22\textwidth}X}
\toprule
Artifact & Use in this work \\
\midrule
MIR benchmark & Primary retrieval benchmark; original chronological train/dev/test split, qrels, and evaluator are preserved. \\
MultiCite-derived supervision & Citation-intent signal used by the released MIR construction; we cite the original source and do not introduce new labels. \\
Stella 1.5B FT anchor & Frozen dense retriever used as the paper-level MIR anchor for M1, M2, and M3. \\
LLM-assisted method blocks & Post-hoc evidence units extracted with DeepSeek-v4pro for diagnosis and evidence-bundle selection only. \\
Data handling & Experiments use publicly released research artifacts, follow their released terms and usage conditions, and do not redistribute the original raw data. \\
\bottomrule
\end{tabularx}
\captionof{table}{Artifacts and reproducibility handling.}
\label{tab:artifact_appendix}
\end{center}

\section{Appendix Case Studies}
We include four representative DeepSeek-v4pro-selected cases from \texttt{cases\_4/}. Each case is presented with its query diagnosis, gap map, and method maturity profile.

\subsection{Case 1: Character-Level Signal Modeling}
\noindent\textbf{Query ID:} \texttt{ABC\_3e344c590b4d5270a29054ac15efa5\_35}. \textbf{Selected reason:} gap=2, partial=0, min coverage/need=0.311. The case concentrates on optimization/efficiency and retrieval/ranking gaps, making it useful for inspecting how the bridge bundle complements sparse methodological coverage.
\begin{figure}[H]
\centering
\includegraphics[width=0.90\textwidth]{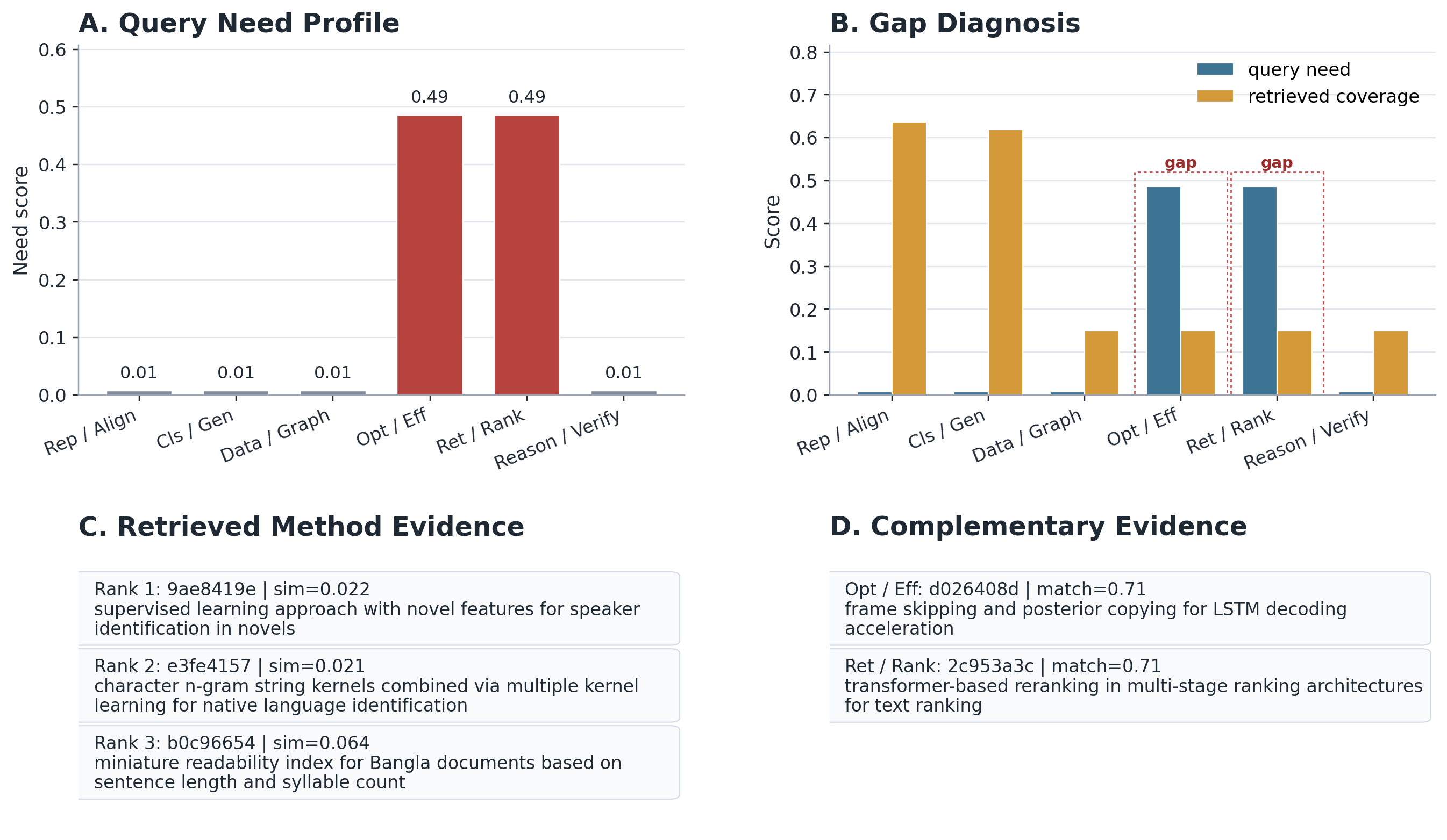}
\vspace{0.2em}
\begin{subfigure}[t]{0.44\textwidth}
\centering
\includegraphics[width=\linewidth]{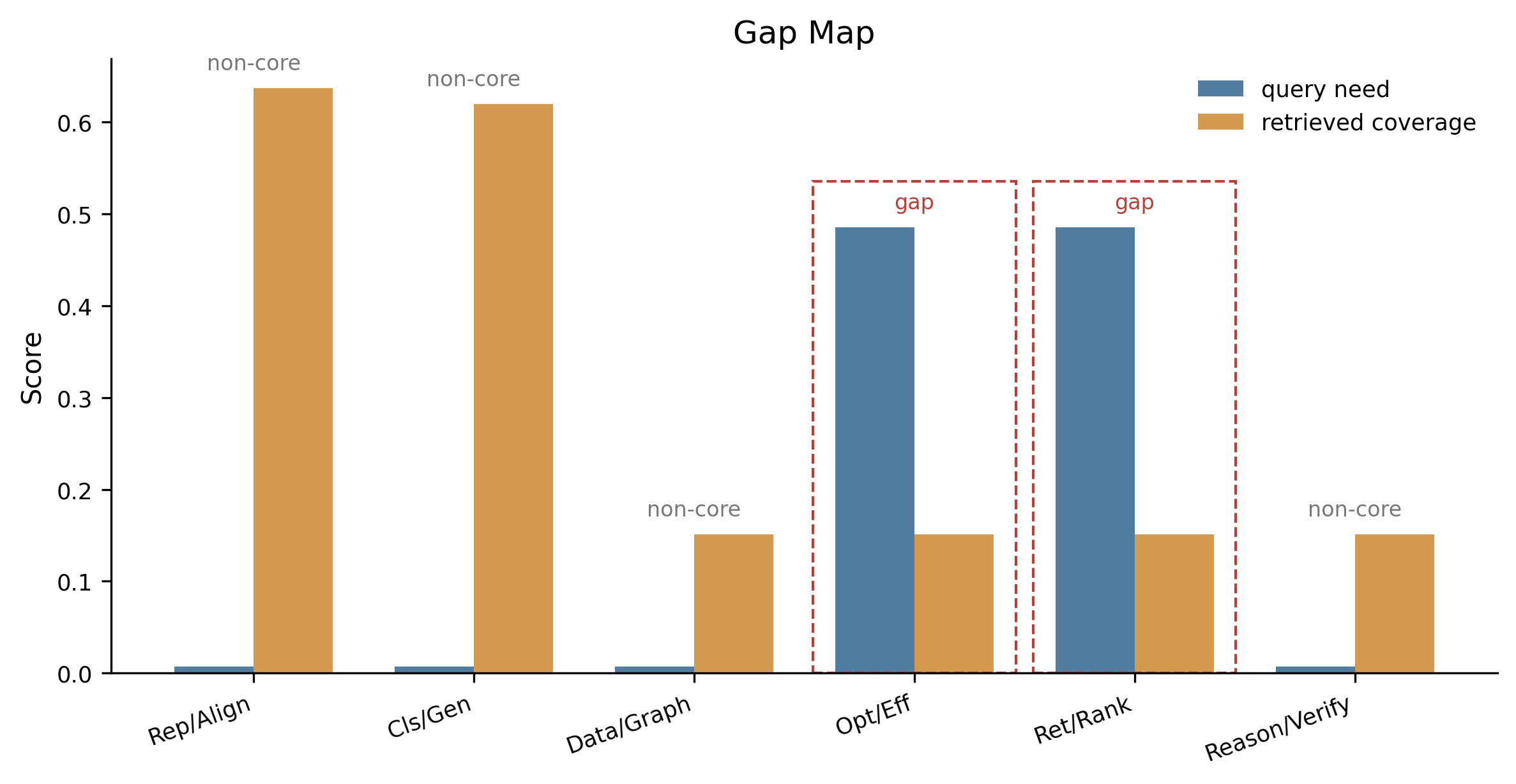}
\caption{Gap map.}
\end{subfigure}\hfill
\begin{subfigure}[t]{0.44\textwidth}
\centering
\includegraphics[width=\linewidth]{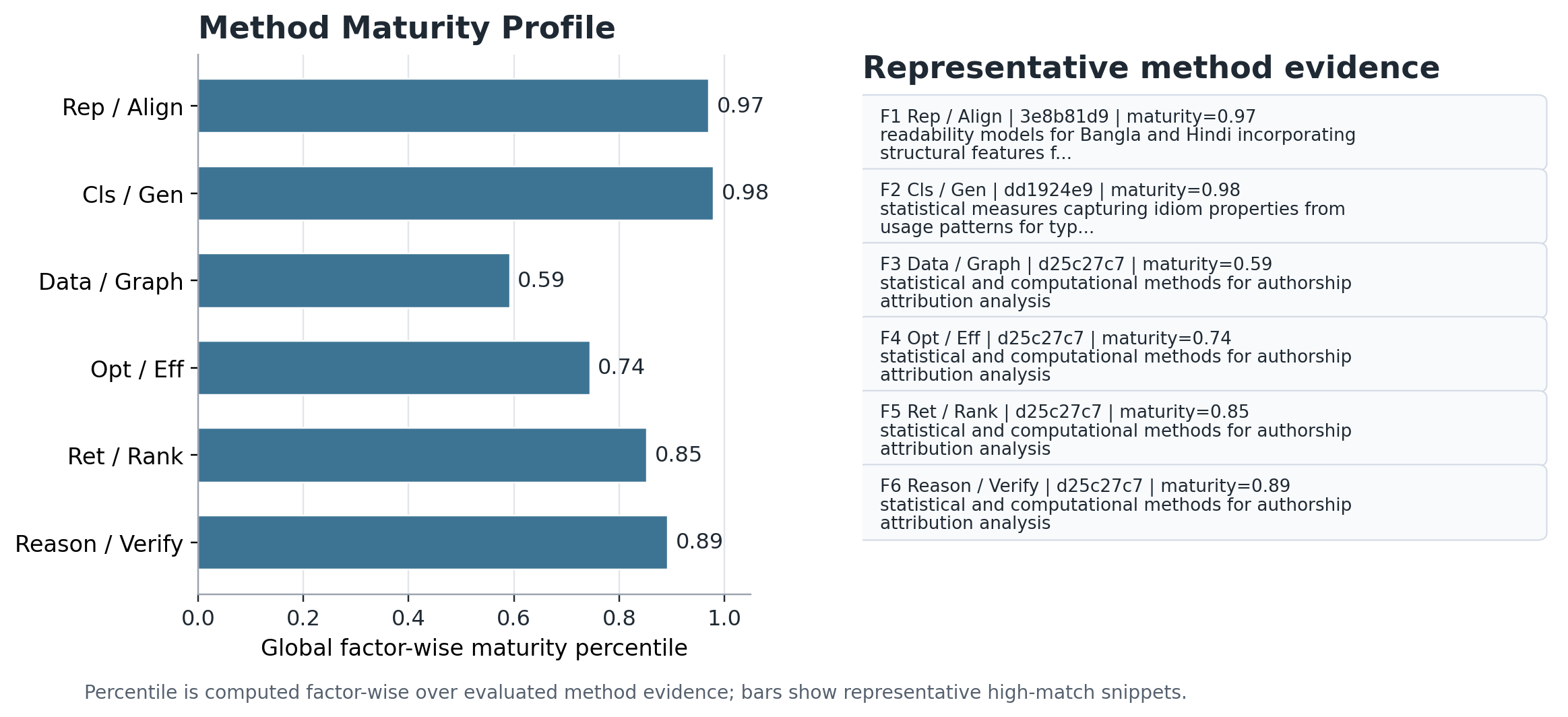}
\caption{Method maturity profile.}
\end{subfigure}
\caption{Appendix case 1. The four-panel summary, gap map, and maturity profile are shown for the selected DeepSeek-v4pro case.}
\label{fig:app_case1}
\end{figure}
\begin{center}
\scriptsize
\setlength{\tabcolsep}{3pt}
\renewcommand{\arraystretch}{1.04}
\begin{tabularx}{0.98\textwidth}{r p{0.13\textwidth} X r}
\toprule
Rank & Paper & DeepSeek method block & Sim. \\
\midrule
1 & \texttt{9ae8419e} & supervised learning approach with novel features for speaker identification in novels & 0.022 \\
2 & \texttt{e3fe4157} & character $n$-gram string kernels combined via multiple kernel learning for native language identification & 0.021 \\
3 & \texttt{b0c96654} & miniature readability index for Bangla documents based on sentence length and syllable count & 0.064 \\
4 & \texttt{d25c27c7} & statistical and computational methods for authorship attribution analysis & 0.122 \\
5 & \texttt{3e8b81d9} & readability models for Bangla and Hindi incorporating structural features from user experiments & 0.091 \\
6 & \texttt{dd1924e9} & statistical measures capturing idiom properties from usage patterns for type-based classification and token identification & 0.041 \\
7 & \texttt{6cff9e93} & document classification trained on parallel corpora for closely related language discrimination & 0.025 \\
7 & \texttt{6cff9e93} & blacklisted word features for emphasizing discriminating characteristics in closely related languages & 0.024 \\
8 & \texttt{cc2a3945} & unsupervised separation of authorial components by exploiting synonym choice differences & 0.000 \\
\bottomrule
\end{tabularx}
\captionof{table}{Retrieved method evidence for appendix case 1. Paper identifiers are abbreviated to the first eight hexadecimal characters.}
\label{tab:app_case1_retrieved}
\end{center}

\begin{center}
\scriptsize
\setlength{\tabcolsep}{3pt}
\renewcommand{\arraystretch}{1.04}
\begin{tabularx}{0.98\textwidth}{p{0.16\textwidth} r p{0.13\textwidth} X p{0.11\textwidth}}
\toprule
Bundle path & \# & Block & Method block & Factor \\
\midrule
Origin & 1 & \texttt{9ae8419e\#b0} & supervised learning approach with novel features for speaker identification in novels & \\
Origin & 2 & \texttt{e3fe4157\#b0} & character $n$-gram string kernels combined via multiple kernel learning for native language identification & \\
Origin & 3 & \texttt{b0c96654\#b0} & miniature readability index for Bangla documents based on sentence length and syllable count & \\
\addlinespace[1pt]
Mutation & 1 & \texttt{9ae8419e\#b0} & supervised learning approach with novel features for speaker identification in novels & \\
Mutation & 2 & \texttt{e3fe4157\#b0} & character $n$-gram string kernels combined via multiple kernel learning for native language identification & \\
Mutation & 3 & \texttt{b0c96654\#b0} & miniature readability index for Bangla documents based on sentence length and syllable count & \\
Mutation & 4 & \texttt{d026408d\#b1} & frame skipping and posterior copying for LSTM decoding acceleration & Opt / Eff \\
Mutation & 5 & \texttt{2c953a3c\#b0} & transformer-based reranking in multi-stage ranking architectures for text ranking & Ret / Rank \\
\addlinespace[1pt]
Crossover & 1 & \texttt{d25c27c7\#b0} & statistical and computational methods for authorship attribution analysis & \\
Crossover & 2 & \texttt{b0c96654\#b0} & miniature readability index for Bangla documents based on sentence length and syllable count & \\
\addlinespace[1pt]
Mutation+crossover & 1 & \texttt{d25c27c7\#b0} & statistical and computational methods for authorship attribution analysis & \\
Mutation+crossover & 2 & \texttt{b0c96654\#b0} & miniature readability index for Bangla documents based on sentence length and syllable count & \\
Mutation+crossover & 3 & \texttt{d026408d\#b1} & frame skipping and posterior copying for LSTM decoding acceleration & Opt / Eff \\
Mutation+crossover & 4 & \texttt{2c953a3c\#b0} & transformer-based reranking in multi-stage ranking architectures for text ranking & Ret / Rank \\
\bottomrule
\end{tabularx}
\captionof{table}{Bundle reconstruction for appendix case 1. Block identifiers are abbreviated; mutation appends complementary blocks, while mutation+crossover combines crossover selections with the mutation additions.}
\label{tab:app_case1_bundles}
\end{center}

\begin{center}
\scriptsize
\setlength{\tabcolsep}{4pt}
\renewcommand{\arraystretch}{1.04}
\begin{tabular}{llrrrrrr}
\toprule
Variant & Path & Rel. & Evid. & Cov. & Coh. & Red. & Fitness \\
\midrule
ours & original & 0.036 & 0.740 & 0.269 & 0.620 & 0.080 & 0.352 \\
ours & mutation & 0.054 & 0.980 & 1.000 & 0.740 & 0.069 & 0.619 \\
ours & crossover & 0.093 & 0.740 & 0.311 & 0.760 & 0.111 & 0.399 \\
ours & mutation+crossover & 0.087 & 0.900 & 1.000 & 0.800 & 0.091 & 0.617 \\
\bottomrule
\end{tabular}
\captionof{table}{M5 path scores for appendix case 1.}
\label{tab:app_case1_m5}
\end{center}

\Needspace{0.88\textheight}
\subsection{Case 2: Sentiment / Opinion Modeling}
\noindent\textbf{Query ID:} \texttt{ABC\_6da7dcbcb7f52f31ec23c8131d438d\_32}. \textbf{Selected reason:} gap=2, partial=0, min coverage/need=0.101. This case has a broader role profile, and the mutation+crossover bundle is the strongest complementary evidence path in the companion M5 table.
\begin{figure}[H]
\centering
\includegraphics[width=0.90\textwidth]{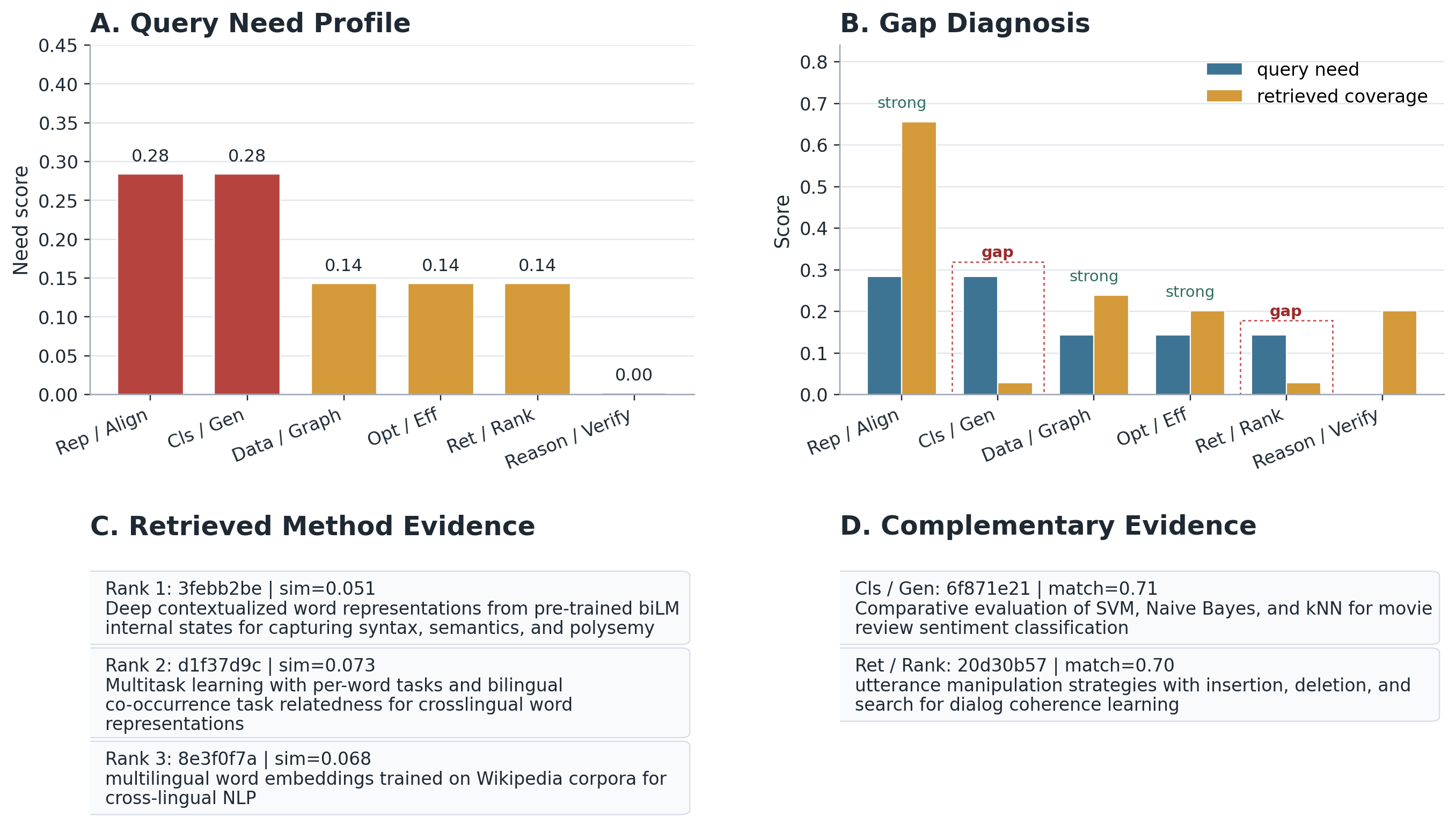}
\vspace{0.2em}
\begin{subfigure}[t]{0.44\textwidth}
\centering
\includegraphics[width=\linewidth]{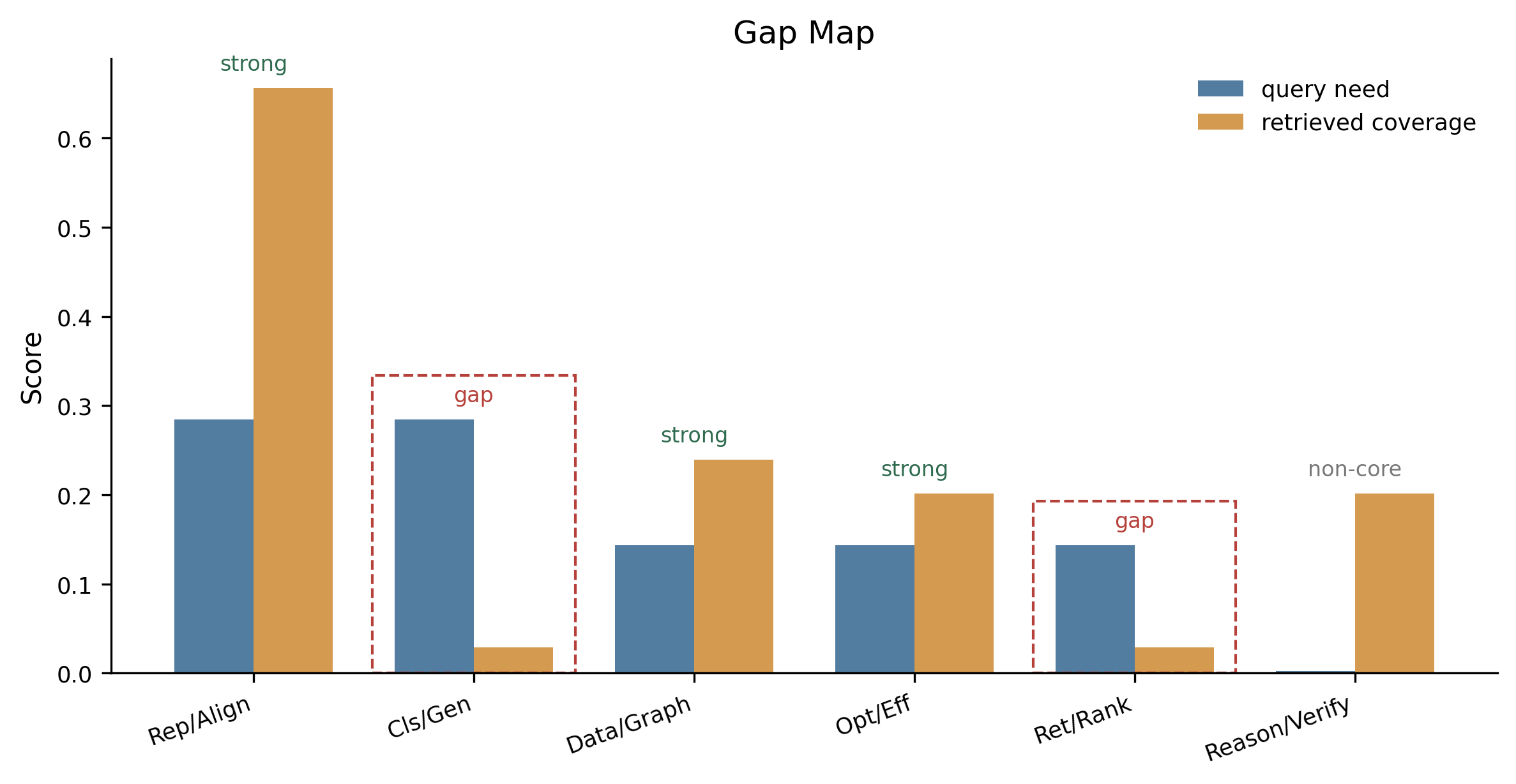}
\caption{Gap map.}
\end{subfigure}\hfill
\begin{subfigure}[t]{0.44\textwidth}
\centering
\includegraphics[width=\linewidth]{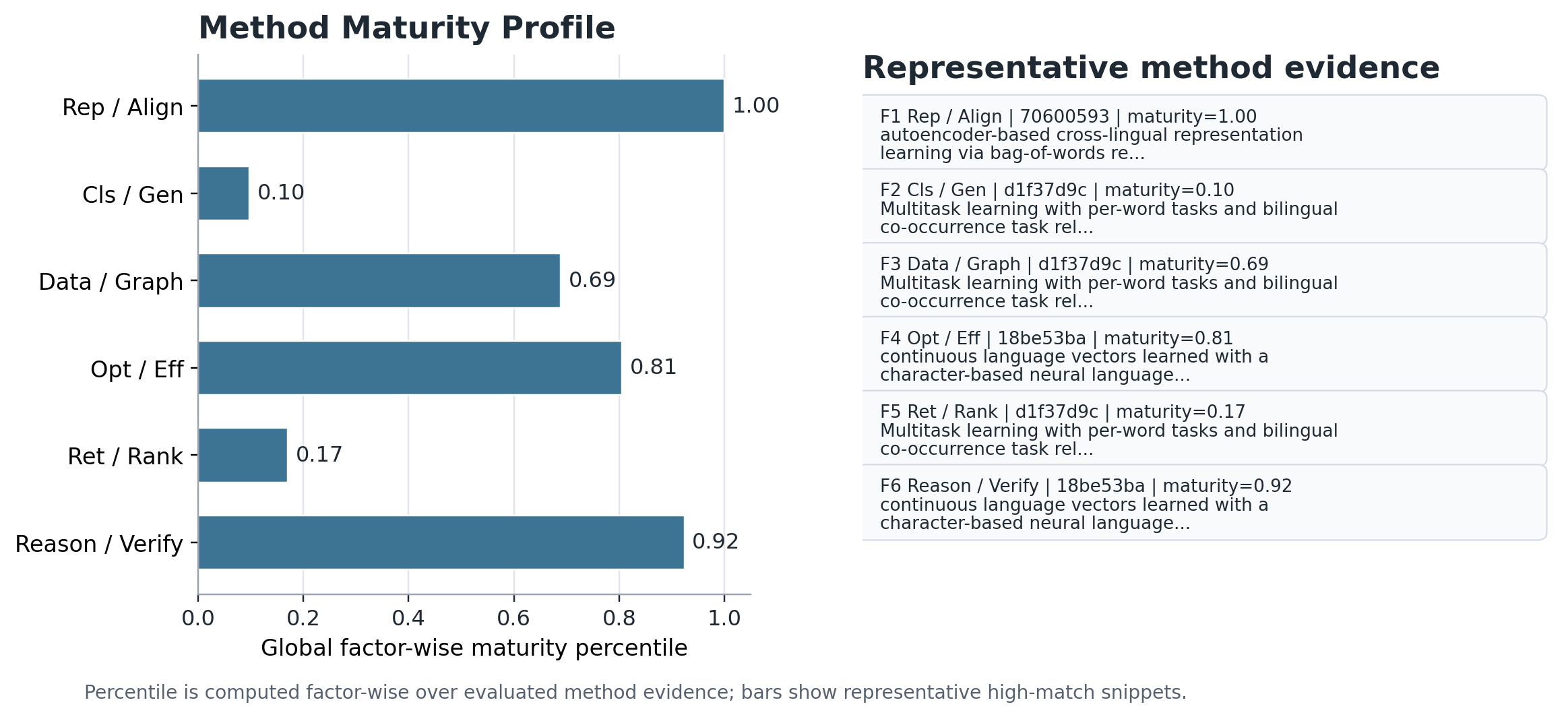}
\caption{Method maturity profile.}
\end{subfigure}
\caption{Appendix case 2. The figure pair summarizes the query need, gap structure, and maturity profile for the selected DeepSeek-v4pro case.}
\label{fig:app_case2}
\end{figure}
\begin{center}
\scriptsize
\setlength{\tabcolsep}{3pt}
\renewcommand{\arraystretch}{1.04}
\begin{tabularx}{0.98\textwidth}{r p{0.13\textwidth} X r}
\toprule
Rank & Paper & DeepSeek method block & Sim. \\
\midrule
1 & \texttt{3febb2be} & Deep contextualized word representations from pre-trained biLM internal states for capturing syntax, semantics, and polysemy & 0.051 \\
2 & \texttt{d1f37d9c} & Multitask learning with per-word tasks and bilingual co-occurrence task relatedness for crosslingual word representations & 0.073 \\
3 & \texttt{8e3f0f7a} & multilingual word embeddings trained on Wikipedia corpora for cross-lingual NLP & 0.068 \\
4 & \texttt{70600593} & autoencoder-based cross-lingual representation learning via bag-of-words reconstruction of aligned sentences without word alignments & 0.052 \\
5 & \texttt{2b669398} & neural architecture learning multi-sense embeddings from local and global document context & 0.051 \\
6 & \texttt{18be53ba} & continuous language vectors learned with a character-based neural language model for improving inference about unseen language varieties & 0.027 \\
7 & \texttt{ac11062f} & comparative study of LSTM, CNN, and self-attention architectures for pre-trained bidirectional language models to analyze contextual representations & 0.060 \\
8 & \texttt{fc303b3b} & BilBOWA: sampled bag-of-words cross-lingual objective to regularize noise-contrastive language models for bilingual word representation learning & 0.073 \\
\bottomrule
\end{tabularx}
\captionof{table}{Retrieved method evidence for appendix case 2. Paper identifiers are abbreviated to the first eight hexadecimal characters.}
\label{tab:app_case2_retrieved}
\end{center}

\begin{center}
\scriptsize
\setlength{\tabcolsep}{3pt}
\renewcommand{\arraystretch}{1.04}
\begin{tabularx}{0.98\textwidth}{p{0.16\textwidth} r p{0.13\textwidth} X p{0.11\textwidth}}
\toprule
Bundle path & \# & Block & Method block & Factor \\
\midrule
Origin & 1 & \texttt{3febb2be\#b0} & Deep contextualized word representations from pre-trained biLM internal states for capturing syntax, semantics, and polysemy & \\
Origin & 2 & \texttt{d1f37d9c\#b0} & Multitask learning with per-word tasks and bilingual co-occurrence task relatedness for crosslingual word representations & \\
Origin & 3 & \texttt{8e3f0f7a\#b0} & multilingual word embeddings trained on Wikipedia corpora for cross-lingual NLP & \\
\addlinespace[1pt]
Mutation & 1 & \texttt{3febb2be\#b0} & Deep contextualized word representations from pre-trained biLM internal states for capturing syntax, semantics, and polysemy & \\
Mutation & 2 & \texttt{d1f37d9c\#b0} & Multitask learning with per-word tasks and bilingual co-occurrence task relatedness for crosslingual word representations & \\
Mutation & 3 & \texttt{8e3f0f7a\#b0} & multilingual word embeddings trained on Wikipedia corpora for cross-lingual NLP & \\
Mutation & 4 & \texttt{6f871e21\#b0} & Comparative evaluation of SVM, Naive Bayes, and kNN for movie review sentiment classification & Cls / Gen \\
Mutation & 5 & \texttt{20d30b57\#b0} & utterance manipulation strategies with insertion, deletion, and search for dialog coherence learning & Ret / Rank \\
\addlinespace[1pt]
Crossover & 1 & \texttt{70600593\#b0} & autoencoder-based cross-lingual representation learning via bag-of-words reconstruction of aligned sentences without word alignments & \\
Crossover & 2 & \texttt{d1f37d9c\#b0} & Multitask learning with per-word tasks and bilingual co-occurrence task relatedness for crosslingual word representations & \\
Crossover & 3 & \texttt{fc303b3b\#b0} & BilBOWA: sampled bag-of-words cross-lingual objective to regularize noise-contrastive language models for bilingual word representation learning & \\
Crossover & 4 & \texttt{18be53ba\#b0} & continuous language vectors learned with a character-based neural language model for improving inference about unseen language varieties & \\
\addlinespace[1pt]
Mutation+crossover & 1 & \texttt{70600593\#b0} & autoencoder-based cross-lingual representation learning via bag-of-words reconstruction of aligned sentences without word alignments & \\
Mutation+crossover & 2 & \texttt{d1f37d9c\#b0} & Multitask learning with per-word tasks and bilingual co-occurrence task relatedness for crosslingual word representations & \\
Mutation+crossover & 3 & \texttt{fc303b3b\#b0} & BilBOWA: sampled bag-of-words cross-lingual objective to regularize noise-contrastive language models for bilingual word representation learning & \\
Mutation+crossover & 4 & \texttt{18be53ba\#b0} & continuous language vectors learned with a character-based neural language model for improving inference about unseen language varieties & \\
Mutation+crossover & 5 & \texttt{6f871e21\#b0} & Comparative evaluation of SVM, Naive Bayes, and kNN for movie review sentiment classification & Cls / Gen \\
Mutation+crossover & 6 & \texttt{20d30b57\#b0} & utterance manipulation strategies with insertion, deletion, and search for dialog coherence learning & Ret / Rank \\
\bottomrule
\end{tabularx}
\captionof{table}{Bundle reconstruction for appendix case 2. Block identifiers are abbreviated; mutation appends complementary blocks, while mutation+crossover combines crossover selections with the mutation additions.}
\label{tab:app_case2_bundles}
\end{center}

\begin{center}
\scriptsize
\setlength{\tabcolsep}{4pt}
\renewcommand{\arraystretch}{1.04}
\begin{tabular}{llrrrrrr}
\toprule
Variant & Path & Rel. & Evid. & Cov. & Coh. & Red. & Fitness \\
\midrule
ours & original & 0.064 & 0.740 & 0.576 & 0.620 & 0.115 & 0.435 \\
ours & mutation & 0.065 & 0.980 & 0.804 & 0.740 & 0.096 & 0.572 \\
ours & crossover & 0.056 & 0.900 & 0.775 & 0.760 & 0.107 & 0.544 \\
ours & mutation+crossover & 0.059 & 1.000 & 1.000 & 0.800 & 0.086 & 0.634 \\
\bottomrule
\end{tabular}
\captionof{table}{M5 path scores for appendix case 2.}
\label{tab:app_case2_m5}
\end{center}

\Needspace{0.88\textheight}
\subsection{Case 3: Negotiation / Automatic Agents}
\noindent\textbf{Query ID:} \texttt{ABC\_b3ef4c176720bdc89d2f73a2560673\_43}. \textbf{Selected reason:} gap=1, partial=1, min coverage/need=0.273. This case is useful because the query is neither fully covered nor fully sparse, so the diagnostic layer must balance bridge completion with evidence selectivity.
\begin{figure}[H]
\centering
\includegraphics[width=0.90\textwidth]{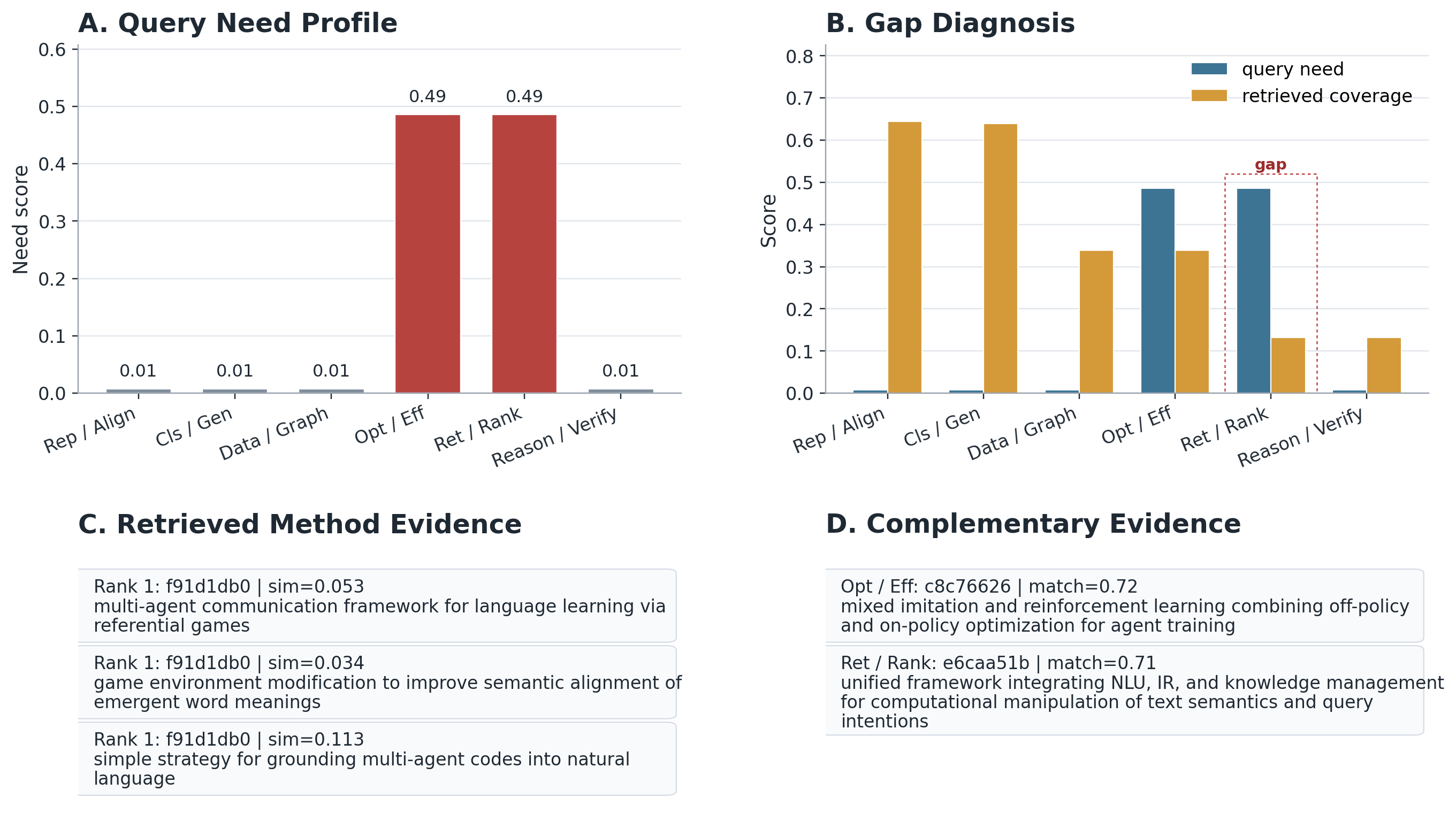}
\vspace{0.2em}
\begin{subfigure}[t]{0.44\textwidth}
\centering
\includegraphics[width=\linewidth]{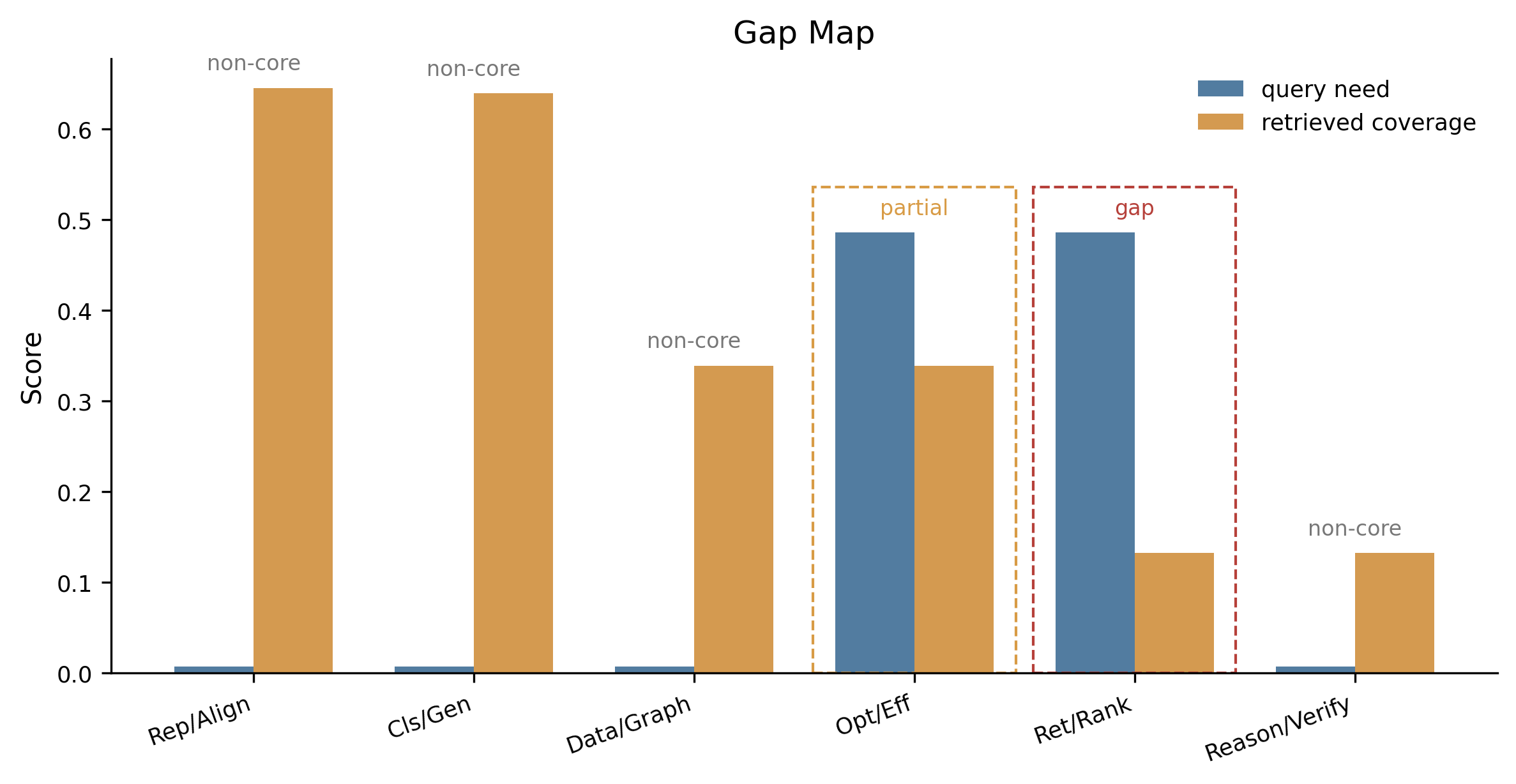}
\caption{Gap map.}
\end{subfigure}\hfill
\begin{subfigure}[t]{0.44\textwidth}
\centering
\includegraphics[width=\linewidth]{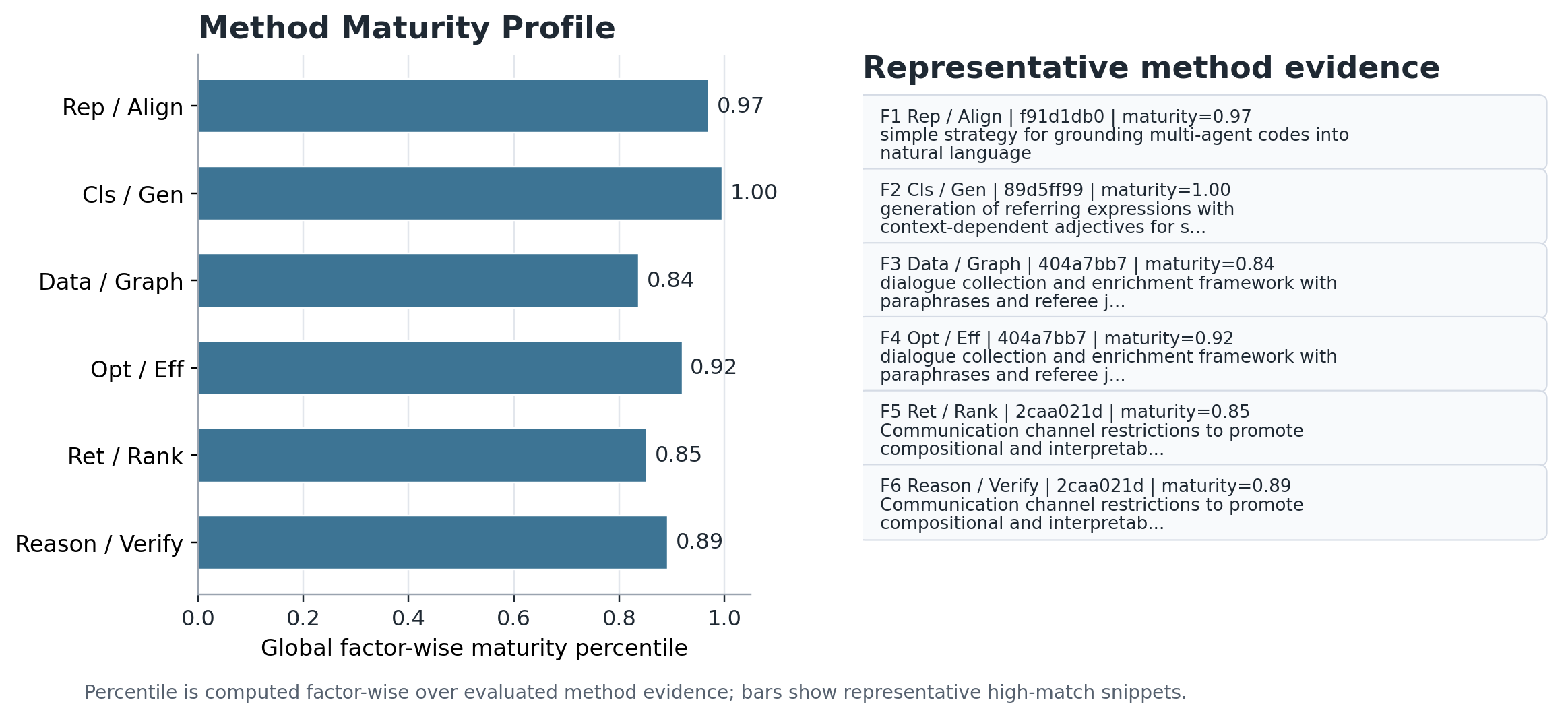}
\caption{Method maturity profile.}
\end{subfigure}
\caption{Appendix case 3. The selected DeepSeek-v4pro case combines partial coverage with a remaining gap, making it suitable for showing complementary evidence composition.}
\label{fig:app_case3}
\end{figure}
\begin{center}
\scriptsize
\setlength{\tabcolsep}{3pt}
\renewcommand{\arraystretch}{1.04}
\begin{tabularx}{0.98\textwidth}{r p{0.13\textwidth} X r}
\toprule
Rank & Paper & DeepSeek method block & Sim. \\
\midrule
1 & \texttt{f91d1db0} & multi-agent communication framework for language learning via referential games & 0.053 \\
1 & \texttt{f91d1db0} & game environment modification to improve semantic alignment of emergent word meanings & 0.034 \\
1 & \texttt{f91d1db0} & simple strategy for grounding multi-agent codes into natural language & 0.113 \\
2 & \texttt{404a7bb7} & dialogue collection and enrichment framework with paraphrases and referee judgments for policy learning data & 0.067 \\
2 & \texttt{404a7bb7} & automatic policy evaluation metric recognizing multiple valid conversational responses per turn & 0.016 \\
3 & \texttt{2caa021d} & Communication channel restrictions to promote compositional and interpretable emergent languages & 0.069 \\
4 & \texttt{4b17662f} & Imitation learning from human demonstrations to transform navigation instructions into plans & 0.050 \\
4 & \texttt{4b17662f} & Learned lexicon to refine inferred plans for navigation instruction following & 0.018 \\
4 & \texttt{4b17662f} & Supervised learning to induce a semantic parser for navigation instructions & 0.035 \\
5 & \texttt{874028f1} & incremental DS-TTR semantic parsing / generation with learned visual classifiers for word meaning grounding & 0.065 \\
5 & \texttt{874028f1} & adaptive dialogue policy for optimizing classifier accuracy vs. tutoring cost trade-off & 0.068 \\
6 & \texttt{1c41e3ef} & rule-based dialogue policy for high performance under perfect NLU assumption & 0.018 \\
6 & \texttt{1c41e3ef} & direct classification combining dialogue policy and NLU for practical advantages & 0.089 \\
7 & \texttt{89d5ff99} & AI planning-based NLG for modeling and manipulating non-linguistic context in situated communication & 0.056 \\
7 & \texttt{89d5ff99} & instruction generation that guides the hearer to a convenient location for referring expressions & 0.036 \\
7 & \texttt{89d5ff99} & generation of referring expressions with context-dependent adjectives for situated descriptions & 0.036 \\
8 & \texttt{d3029c06} & Neural Belief Tracking framework leveraging pre-trained word vectors to learn distributed representations for scalable state tracking & 0.030 \\
\bottomrule
\end{tabularx}
\captionof{table}{Retrieved method evidence for appendix case 3. Paper identifiers are abbreviated to the first eight hexadecimal characters.}
\label{tab:app_case3_retrieved}
\end{center}

\begin{center}
\scriptsize
\setlength{\tabcolsep}{3pt}
\renewcommand{\arraystretch}{1.04}
\begin{tabularx}{0.98\textwidth}{p{0.16\textwidth} r p{0.13\textwidth} X p{0.11\textwidth}}
\toprule
Bundle path & \# & Block & Method block & Factor \\
\midrule
Origin & 1 & \texttt{f91d1db0\#b0} & multi-agent communication framework for language learning via referential games & \\
Origin & 2 & \texttt{f91d1db0\#b1} & game environment modification to improve semantic alignment of emergent word meanings & \\
Origin & 3 & \texttt{f91d1db0\#b2} & simple strategy for grounding multi-agent codes into natural language & \\
\addlinespace[1pt]
Mutation & 1 & \texttt{f91d1db0\#b0} & multi-agent communication framework for language learning via referential games & \\
Mutation & 2 & \texttt{f91d1db0\#b1} & game environment modification to improve semantic alignment of emergent word meanings & \\
Mutation & 3 & \texttt{f91d1db0\#b2} & simple strategy for grounding multi-agent codes into natural language & \\
Mutation & 4 & \texttt{c8c76626\#b0} & mixed imitation and reinforcement learning combining off-policy and on-policy optimization for agent training & Opt / Eff \\
Mutation & 5 & \texttt{e6caa51b\#b0} & unified framework integrating NLU, IR, and knowledge management for computational manipulation of text semantics and query intentions & Ret / Rank \\
\addlinespace[1pt]
Crossover & 1 & \texttt{404a7bb7\#b0} & dialogue collection and enrichment framework with paraphrases and referee judgments for policy learning data & \\
Crossover & 2 & \texttt{2caa021d\#b0} & Communication channel restrictions to promote compositional and interpretable emergent languages & \\
\addlinespace[1pt]
Mutation+crossover & 1 & \texttt{404a7bb7\#b0} & dialogue collection and enrichment framework with paraphrases and referee judgments for policy learning data & \\
Mutation+crossover & 2 & \texttt{2caa021d\#b0} & Communication channel restrictions to promote compositional and interpretable emergent languages & \\
Mutation+crossover & 3 & \texttt{c8c76626\#b0} & mixed imitation and reinforcement learning combining off-policy and on-policy optimization for agent training & Opt / Eff \\
Mutation+crossover & 4 & \texttt{e6caa51b\#b0} & unified framework integrating NLU, IR, and knowledge management for computational manipulation of text semantics and query intentions & Ret / Rank \\
\bottomrule
\end{tabularx}
\captionof{table}{Bundle reconstruction for appendix case 3. Block identifiers are abbreviated; mutation appends complementary blocks, while mutation+crossover combines crossover selections with the mutation additions.}
\label{tab:app_case3_bundles}
\end{center}

\begin{center}
\scriptsize
\setlength{\tabcolsep}{4pt}
\renewcommand{\arraystretch}{1.04}
\begin{tabular}{llrrrrrr}
\toprule
Variant & Path & Rel. & Evid. & Cov. & Coh. & Red. & Fitness \\
\midrule
ours & original & 0.067 & 0.740 & 0.261 & 0.620 & 0.067 & 0.360 \\
ours & mutation & 0.083 & 0.980 & 1.000 & 0.740 & 0.058 & 0.628 \\
ours & crossover & 0.068 & 0.740 & 0.485 & 0.760 & 0.048 & 0.438 \\
ours & mutation+crossover & 0.088 & 0.900 & 1.000 & 0.800 & 0.081 & 0.617 \\
\bottomrule
\end{tabular}
\captionof{table}{M5 path scores for appendix case 3.}
\label{tab:app_case3_m5}
\end{center}

\Needspace{0.88\textheight}
\subsection{Case 4: Sentiment / Opinion Modeling Additional Trace}
\noindent\textbf{Query ID:} \texttt{ABC\_6da7dcbcb7f52f31ec23c8131d438d\_32}. \textbf{Selected reason:} gap=2, partial=0, min coverage/need=0.101. This additional trace uses the same sentiment/opinion modeling query family and shows the corresponding query diagnosis, gap map, and maturity profile from the four-case package.
\begin{figure}[H]
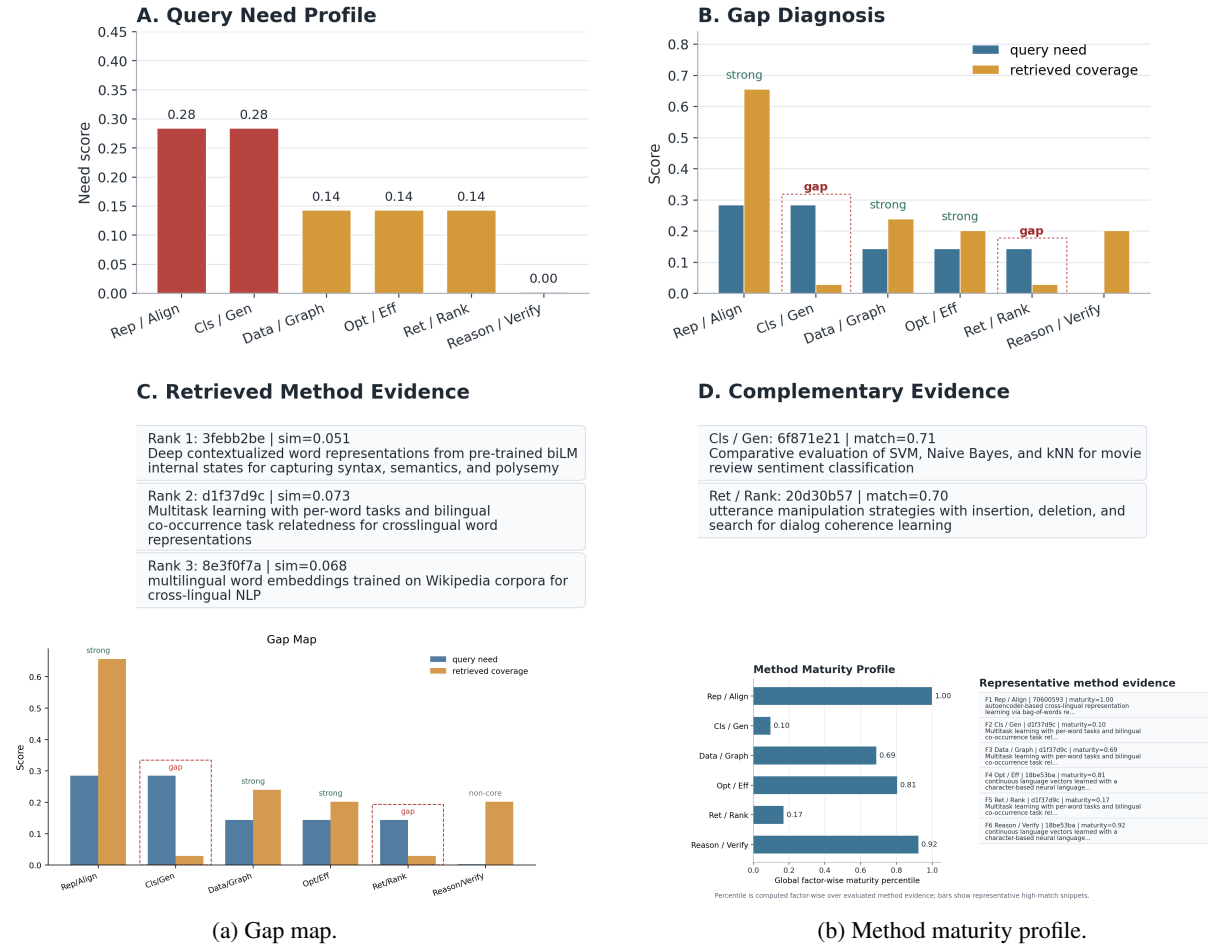

\centering
\includegraphics[width=0.90\textwidth]{cases_4/case_04_ABC_6da7dcbcb7f52f31ec23c8131d438d_32/four_panel_case_study.png}
\vspace{0.2em}
\begin{subfigure}[t]{0.44\textwidth}
\centering
\includegraphics[width=\linewidth]{cases_4/case_04_ABC_6da7dcbcb7f52f31ec23c8131d438d_32/gap_map.png}
\caption{Gap map.}
\end{subfigure}\hfill
\begin{subfigure}[t]{0.44\textwidth}
\centering
\includegraphics[width=\linewidth]{cases_4/case_04_ABC_6da7dcbcb7f52f31ec23c8131d438d_32/method_maturity_profile.png}
\caption{Method maturity profile.}
\end{subfigure}
\caption{Appendix case 4. The four-panel summary, gap map, and maturity profile are shown for the additional DeepSeek-v4pro trace.}
\label{fig:app_case4}
\end{figure}
\begin{center}
\scriptsize
\setlength{\tabcolsep}{3pt}
\renewcommand{\arraystretch}{1.04}
\begin{tabularx}{0.98\textwidth}{r p{0.13\textwidth} X r}
\toprule
Rank & Paper & DeepSeek method block & Sim. \\
\midrule
1 & \texttt{3febb2be} & Deep contextualized word representations from pre-trained biLM internal states for capturing syntax, semantics, and polysemy & 0.051 \\
2 & \texttt{d1f37d9c} & Multitask learning with per-word tasks and bilingual co-occurrence task relatedness for crosslingual word representations & 0.073 \\
3 & \texttt{8e3f0f7a} & multilingual word embeddings trained on Wikipedia corpora for cross-lingual NLP & 0.068 \\
4 & \texttt{70600593} & autoencoder-based cross-lingual representation learning via bag-of-words reconstruction of aligned sentences without word alignments & 0.052 \\
5 & \texttt{2b669398} & neural architecture learning multi-sense embeddings from local and global document context & 0.051 \\
6 & \texttt{18be53ba} & continuous language vectors learned with a character-based neural language model for improving inference about unseen language varieties & 0.027 \\
7 & \texttt{ac11062f} & comparative study of LSTM, CNN, and self-attention architectures for pre-trained bidirectional language models to analyze contextual representations & 0.060 \\
8 & \texttt{fc303b3b} & BilBOWA: sampled bag-of-words cross-lingual objective to regularize noise-contrastive language models for bilingual word representation learning & 0.073 \\
\bottomrule
\end{tabularx}
\captionof{table}{Retrieved method evidence for appendix case 4. Paper identifiers are abbreviated to the first eight hexadecimal characters.}
\label{tab:app_case4_retrieved}
\end{center}

\begin{center}
\scriptsize
\setlength{\tabcolsep}{3pt}
\renewcommand{\arraystretch}{1.04}
\begin{tabularx}{0.98\textwidth}{p{0.16\textwidth} r p{0.13\textwidth} X p{0.11\textwidth}}
\toprule
Bundle path & \# & Block & Method block & Factor \\
\midrule
Origin & 1 & \texttt{3febb2be\#b0} & Deep contextualized word representations from pre-trained biLM internal states for capturing syntax, semantics, and polysemy & \\
Origin & 2 & \texttt{d1f37d9c\#b0} & Multitask learning with per-word tasks and bilingual co-occurrence task relatedness for crosslingual word representations & \\
Origin & 3 & \texttt{8e3f0f7a\#b0} & multilingual word embeddings trained on Wikipedia corpora for cross-lingual NLP & \\
\addlinespace[1pt]
Mutation & 1 & \texttt{3febb2be\#b0} & Deep contextualized word representations from pre-trained biLM internal states for capturing syntax, semantics, and polysemy & \\
Mutation & 2 & \texttt{d1f37d9c\#b0} & Multitask learning with per-word tasks and bilingual co-occurrence task relatedness for crosslingual word representations & \\
Mutation & 3 & \texttt{8e3f0f7a\#b0} & multilingual word embeddings trained on Wikipedia corpora for cross-lingual NLP & \\
Mutation & 4 & \texttt{6f871e21\#b0} & Comparative evaluation of SVM, Naive Bayes, and kNN for movie review sentiment classification & Cls / Gen \\
Mutation & 5 & \texttt{20d30b57\#b0} & utterance manipulation strategies with insertion, deletion, and search for dialog coherence learning & Ret / Rank \\
\addlinespace[1pt]
Crossover & 1 & \texttt{70600593\#b0} & autoencoder-based cross-lingual representation learning via bag-of-words reconstruction of aligned sentences without word alignments & \\
Crossover & 2 & \texttt{d1f37d9c\#b0} & Multitask learning with per-word tasks and bilingual co-occurrence task relatedness for crosslingual word representations & \\
Crossover & 3 & \texttt{fc303b3b\#b0} & BilBOWA: sampled bag-of-words cross-lingual objective to regularize noise-contrastive language models for bilingual word representation learning & \\
Crossover & 4 & \texttt{18be53ba\#b0} & continuous language vectors learned with a character-based neural language model for improving inference about unseen language varieties & \\
\addlinespace[1pt]
Mutation+crossover & 1 & \texttt{70600593\#b0} & autoencoder-based cross-lingual representation learning via bag-of-words reconstruction of aligned sentences without word alignments & \\
Mutation+crossover & 2 & \texttt{d1f37d9c\#b0} & Multitask learning with per-word tasks and bilingual co-occurrence task relatedness for crosslingual word representations & \\
Mutation+crossover & 3 & \texttt{fc303b3b\#b0} & BilBOWA: sampled bag-of-words cross-lingual objective to regularize noise-contrastive language models for bilingual word representation learning & \\
Mutation+crossover & 4 & \texttt{18be53ba\#b0} & continuous language vectors learned with a character-based neural language model for improving inference about unseen language varieties & \\
Mutation+crossover & 5 & \texttt{6f871e21\#b0} & Comparative evaluation of SVM, Naive Bayes, and kNN for movie review sentiment classification & Cls / Gen \\
Mutation+crossover & 6 & \texttt{20d30b57\#b0} & utterance manipulation strategies with insertion, deletion, and search for dialog coherence learning & Ret / Rank \\
\bottomrule
\end{tabularx}
\captionof{table}{Bundle reconstruction for appendix case 4. Block identifiers are abbreviated; mutation appends complementary blocks, while mutation+crossover combines crossover selections with the mutation additions.}
\label{tab:app_case4_bundles}
\end{center}

\begin{center}
\scriptsize
\setlength{\tabcolsep}{4pt}
\renewcommand{\arraystretch}{1.04}
\begin{tabular}{llrrrrrr}
\toprule
Variant & Path & Rel. & Evid. & Cov. & Coh. & Red. & Fitness \\
\midrule
ours & original & 0.064 & 0.740 & 0.576 & 0.620 & 0.115 & 0.435 \\
ours & mutation & 0.065 & 0.980 & 0.804 & 0.740 & 0.096 & 0.572 \\
ours & crossover & 0.056 & 0.900 & 0.775 & 0.760 & 0.107 & 0.544 \\
ours & mutation+crossover & 0.059 & 1.000 & 1.000 & 0.800 & 0.086 & 0.634 \\
\bottomrule
\end{tabular}
\captionof{table}{M5 path scores for appendix case 4.}
\label{tab:app_case4_m5}
\end{center}

\section{Implementation and Compute Details}
\begin{center}
\centering
\small
\begin{tabularx}{0.96\textwidth}{p{0.24\textwidth}X}
\toprule
Item & Value \\
\midrule
Dense anchor & MIR Stella 1.5B FT extended dense retriever \\
Anchor status & frozen in the M1, M2, and M3 bridge variants \\
Trained modules & hyperbolic bridge projection and scoring layers; M3 factor bridge with role, sparsity, diversity, and cone losses \\
GPU & NVIDIA RTX PRO 6000 Blackwell Workstation Edition \\
Device & cuda:0 with CUDA\_VISIBLE\_DEVICES=0 \\
Diagnostic run window & 2026-05-23 21:27:32--21:33:16 \\
Approx. diagnostic compute & 5.7 minutes, or about 0.096 GPU-hour on one GPU \\
Runtime memory snapshot & 4716 / 97887 MB on GPU 0 at diagnostic completion \\
\bottomrule
\end{tabularx}
\captionof{table}{Implementation and compute summary from the recorded M3 diagnostic run.}
\label{tab:compute_appendix}
\end{center}

\section{Training Objective Definitions}
For a query $q$, positive candidate $d^+$, negative set $\mathcal{N}(q)$, bridge score $s(q,d)$, temperature $\tau$, normalized factor activation $a(q,d)$, role label $r$, factor directions $u_i$, and cone aperture $\psi(z_q)$, the M3 training objective is
\begin{equation}
\begin{aligned}
L ={}& L_{\mathrm{ret}}
 + \lambda_{\mathrm{role}} L_{\mathrm{role}}
 + \lambda_{\mathrm{sparse}} L_{\mathrm{sparse}} \\
&+ \lambda_{\mathrm{div}} L_{\mathrm{div}}
 + \lambda_{\mathrm{cone}} L_{\mathrm{cone}} .
\end{aligned}
\end{equation}
The retrieval component is the contrastive loss
\begin{equation}
\begin{aligned}
L_{\mathrm{ret}} ={}&
-\log
\frac{\exp(s(q,d^+)/\tau)}
{\exp(s(q,d^+)/\tau)
 + \sum_{d^-\in\mathcal{N}(q)}
 \exp(s(q,d^-)/\tau)} .
\end{aligned}
\end{equation}

\begin{center}
\centering
\small
\begin{tabularx}{0.96\textwidth}{p{0.18\textwidth}X}
\toprule
Term & Definition \\
\midrule
$L_{\mathrm{ret}}$ & Contrastive retrieval loss over one positive candidate and sampled negatives. \\
$L_{\mathrm{role}}$ & Role-alignment cross entropy, $-\log p(r\mid q,d)$, where $p$ is predicted from factor activations. \\
$L_{\mathrm{sparse}}$ & Activation entropy penalty, $H(a(q,d))$, encouraging concentrated factor use. \\
$L_{\mathrm{div}}$ & Factor-diversity penalty, $\sum_{i<j}(\hat{u}_i^\top \hat{u}_j)^2$, discouraging duplicate factor directions. \\
$L_{\mathrm{cone}}$ & Directional compatibility penalty, $[\angle(z_d,z_q)-\psi(z_q)]_+$, applied when a candidate falls outside the query-induced cone. \\
\bottomrule
\end{tabularx}
\captionof{table}{Definitions of the M3 objective terms.}
\label{tab:objective_appendix}
\end{center}

\section{Training Hyperparameters}
\begin{center}
\centering
\small
\begin{tabularx}{0.96\textwidth}{p{0.24\textwidth}X}
\toprule
Hyperparameter & Value \\
\midrule
Batch size & 512 \\
Epochs & 5 \\
Learning rate & $5\times10^{-5}$ \\
Selected factors & $K=6$ \\
$\lambda_{\mathrm{role}}$ & 0.10 \\
$\lambda_{\mathrm{sparse}}$ & 0.10 \\
$\lambda_{\mathrm{div}}$ & 0.05 \\
$\lambda_{\mathrm{cone}}$ & 0.01 \\
Random seeds & 2026052404 and 2026052430 for K=6 diagnostics \\
\bottomrule
\end{tabularx}
\captionof{table}{Hyperparameters used for the selected M3-VisFactor diagnostic backbone.}
\label{tab:hyperparams_appendix}
\end{center}

\section{Software Environment}
\begin{center}
\centering
\small
\begin{tabularx}{0.96\textwidth}{p{0.24\textwidth}X}
\toprule
Item & Value \\
\midrule
Operating system & Ubuntu 22.04.5 LTS \\
PyTorch & 2.11.0+cu130 for diagnostics \\
CUDA & 13.0 for diagnostics \\
GPU driver & 580.159.03 \\
Evaluation protocol & released MIR qrels, chronological splits, and evaluator preserved \\
Random seeds & 2026052404 and 2026052430 for K=6 diagnostics \\
\bottomrule
\end{tabularx}
\captionof{table}{Software environment and evaluation protocol summary.}
\label{tab:software_appendix}
\end{center}

\end{document}